\documentclass[aps,prb,twocolumn,floatfix,10pt,eqsecnum,superscriptaddress]{revtex4-2}

\usepackage[T1]{fontenc}
\usepackage{amsmath}
\usepackage{amssymb}
\usepackage{amsfonts}
\usepackage{graphicx}
\usepackage{color}
\usepackage{xcolor}
\usepackage{bbm}
\usepackage{mathptmx}
\usepackage[breaklinks=true,colorlinks=true,linkcolor=blue,urlcolor=blue,citecolor=blue]{hyperref}

\DeclareMathAlphabet{\mathcal}{OMS}{cmsy}{m}{n}

\newcommand{\debug}[1]{{\iffalse\color{ultramarine} \textbf{#1}\fi}}

\begin{document}

\title{Conformal approach to physics simulations for thin curved 3D membranes}
\author{Igor \surname{Bogush}}
\email{igori.bogus@tu-braunschweig.de}
\affiliation{Cryogenic Quantum Electronics, Laboratory for Emerging Nanometrology (LENA), Institute for Electrical Measurement Science and Fundamental Electrical Engineering, Technische Universit\"at Braunschweig, Hans-Sommer-Str. 66, 38106 Braunschweig, Germany}

\author{Vladimir M. \surname{Fomin}}
\affiliation{Institute for Emerging Electronic Technologies, Leibniz IFW Dresden,
Helmholtzstra\ss e 20, 01069 Dresden, Germany}
\affiliation{Faculty of Physics and Engineering, Moldova State University, str. Alexei Mateevici 60, MD-2009 Chişinău, Republic of Moldova}

\author{Oleksandr V. \surname{Dobrovolskiy}}
\affiliation{Cryogenic Quantum Electronics, Laboratory for Emerging Nanometrology (LENA), Institute for Electrical Measurement Science and Fundamental Electrical Engineering, Technische Universit\"at Braunschweig, Hans-Sommer-Str. 66, 38106 Braunschweig, Germany}

\begin{abstract}
Three-dimensional nanoarchitectures are widely used across various areas of physics, including spintronics, photonics, and superconductivity. In this regard, thin curved 3D membranes are especially interesting for applications in nano- and optoelectronics, sensorics, and information processing, making physics simulations in complex 3D geometries indispensable for unveiling new physical phenomena and the development of devices. Here, we present a general-purpose approach to physics simulations for thin curved 3D membranes, that allows for performing simulations using finite difference methods instead of meshless methods or methods with irregular meshes. The approach utilizes a numerical conformal mapping of the initial surface to a flat domain and is based on the uniformization theorem stating that any simply-connected Riemann surface is conformally equivalent to an open unit disk, a complex plane, or a Riemann sphere. We reveal that for many physical problems involving the Laplace operator and divergence, a flat-domain formulation of the initial problem only requires a modification of the equations of motion and the boundary conditions by including a conformal factor and the mean/Gaussian curvatures. We demonstrate the method's capabilities for case studies of the Schr\"{o}dinger equation for a charged particle in static electric and magnetic fields for 3D geometries, including C-shaped and ring-shaped structures, as well as for the time-dependent Ginzburg-Landau equation.
\end{abstract}
 
\maketitle
\section{Introduction}
Three-dimensional (3D) nanoarchitectures are subject of extensive theoretical and experimental studies with prospects for applications in nano- and optoelectronics, quantum optics, sensorics, and information processing \cite{Thurmer10,Fomin2012NL,Loe19acs,shani2020dna,Fomin21book,Fomin2022APL,Cordoba24apl,Bog24prb}. One of the interesting classes of 3D nanoarchitectures are thin films or membranes (thin material layers without a substrate or when the substrate is irrelevant) of curved geometry \cite{Mak21adm}. Such structures can be fabricated using established methods which range from focused ion/electron beam induced deposition \cite{cordoba2019three,porrati2019crystalline,zhakina2024vortex,fernandez2020writing} over lithography \cite{erbas2024combining} and origami/kirigami techniques \cite{chen2020kirigami} to self-organization driven by mechanical strain relaxation \cite{schmidt2001thin}. The complex geometry affects the electronic \cite{fom07prb,gen22nat}, photonic \cite{ma16nat}, phononic \cite{bal12mto}, magnetic \cite{mak22spr,Gub24pcm}, and superconducting \cite{Loe19acs} properties, and thus requires general-purpose approaches to physics simulations for curved 3D membranes.

In general, the complex 3D geometry affects the system dynamics in several ways. First, this is a nonuniform dilation and contraction of the sample under consideration. For example, by virtue of the sample contraction, the local current density can be increased (current-crowding effect \cite{Bez22prb}), which can lead to a destruction of superconductivity. Second, this is a curvature-induced potential \cite{da1981quantum,jen1971ann}
associated with nontrivial geometry. The third effect lies in the appearance of a non-trivial profile of the normal and/or tangential components of external fields \cite{Fomin21book}. There are also further effects caused by the deformations of the crystal lattice \cite{schmidt2001thin}. However, these effects are beyond the scope of this work and they will be considered elsewhere. 

The system dynamics in physics simulations is usually described by (a system of) partial differential equations (PDEs) \cite{Fomin21book}. The choice of a numerical method suitable for physics simulations depends on how effectively it can implement the geometry and equations of the considered system, ensuring high computational efficiency and simplicity of implementation \cite{Mak21adm}. The most common numerical methods for solving PDEs include finite difference methods (FDMs), finite element methods (FEMs), and finite volume methods (FVMs). The FDMs utilize rectangular grids, are simple to implement, but are difficult to apply to complex geometries. In contrast, the FEMs use irregular meshes, are harder to implement, but they can be applied to a much wider range of complex geometries. Additionally, the FEMs require solving sparse systems of linear equations at each time step, unlike FDMs. Solving the systems of equations reduces the computational efficiency of the FEMs. A compromise between the FDMs and FEMs is offered by the FVMs which use an irregular mesh, are usually simpler to implement than the FEMs, can be applied to complex geometries, and do not require solving the systems of equations for the next time step. However, while the FDMs and FEMs can simulate a wide range of PDEs, the FVMs are better suited to equations where each term involving a differential operator can be represented as a divergence.

The computational efficiency also depends on the hardware used for simulations. Over the past two decades, the use of graphics processing units (GPUs) for numerical simulations has increased significantly. The release of parallel computing platforms exploiting memory coalescing has made algorithms with regular grids, such as the FDMs, more performant for GPU-based simulations. Thus, expanding the applicability of the FDMs to more complex geometries could preserve both efficiency and simplicity. However, a direct application of the FDMs to differential operators in an arbitrary coordinate system can increase the complexity of the numerical scheme and eventually decrease accuracy, due to such factors as highly non-uniform grid sizes, small angles between the basis vectors, and coordinate singularities. Accordingly, a general-purpose approach, which is free of these drawbacks, is required for physics simulations of curved 3D membranes.

Here, we present an approach to physics simulations for thin membranes with non-planar geometry within the FDM paradigm. The technique is applicable to curved 3D membranes in the limit of infinitely small thickness. The method utilizes a 2D regular grid, which can be extended to a 3D regular grid to capture nonlinear effects and the impact of external fields on the system dynamics across the membrane thickness. The approach is based on the uniformization theorem, stating that any simply-connected Riemann surface is conformally equivalent to an open unit disk (a disk with unit radius without points on its boundary), a complex plane, or a Riemann sphere \cite{riemann1851,poincare1908,koebe1907}. That is, for any simply-connected surface representing a physical membrane, there is always a conformal transformation that translates the surface to a finite flat domain. Consequently, the initial surface can be obtained by applying an inverse conformal transformation to a flat domain.

The action of the conformal transformations on the Laplace operator and divergence results in the inclusion of a multiplicative conformal factor. In addition, reducing a physical membrane with finite thickness to a 2D mathematical surface requires the inclusion of a geometric potential that depends on the mean and Gaussian curvatures \cite{da1981quantum,jen1971ann}. Accordingly, for many problems involving the Laplace operator and divergence, a flat-domain formulation of the initial problem with nonplanar geometry only requires a simple modification of equations and boundary conditions by including a conformal factor and the mean/Gaussian curvatures. The developed method allows for the modeling of curved 3D membranes, preserving the high performance of FDMs on GPUs. We demonstrate the capabilities of the modeling procedure for a case study of the Schr\"{o}dinger equation for a charged particle in static uniform electric and magnetic fields for a series of geometries, including a C-shaped and ring-shaped geometries resembling a crater represented by a 2D grid. An extension of this procedure to 3D grids is demonstrated for a case study of the time-dependent Ginzburg-Landau equation (TDGL).

\section{Model}\label{sModel}
\begin{figure*}[t!]
    \centering
    \includegraphics[width=0.8\linewidth]{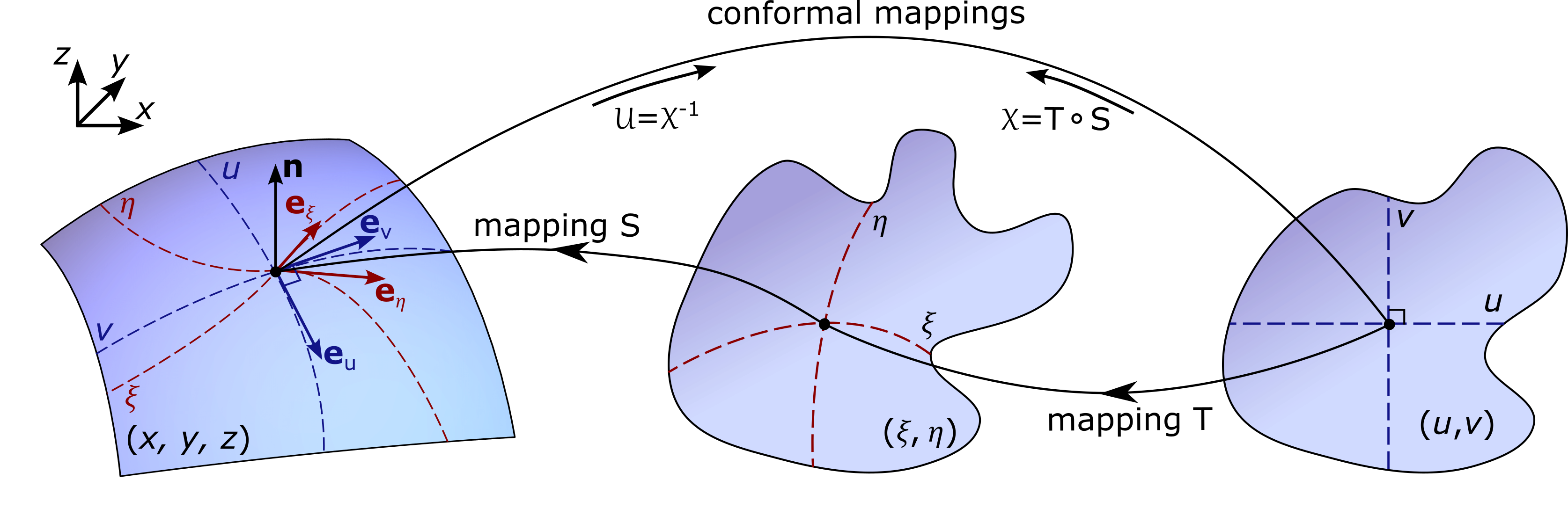}
    \caption{Mappings $S,\,T,\,\mathcal{X},\,\mathcal{U}$ between points in the coordinates $(u, v)$, $(\xi, \eta)$, and $(x, y, z)$. Dashed lines denote the coordinate lines going through the corresponding points.}
    \label{fig:1}
\end{figure*}

\subsection{Qualitative consideration}

We refer to a membrane as a physical object with a finite thickness, which is small compared to its length and width, while a surface is understood as a mathematical construct without thickness. Our goal is to reduce the dynamics of a physical system, described by a scalar function $\psi$, from a 3D membrane to a 2D surface and reformulate the problem of modeling this physical system in terms of flat differential operators. To this end, we first describe the geometry of the membrane and consider the differential operators that are commonly present in various problems of mathematical physics represented by partial differential equations. Among these are the divergence and the gauge-invariant Laplace-Beltrami operator being a generalization of the Laplacian to complex geometries. Next, we reduce the operators to the limit of vanishing thickness and reformulate the initial equations in arbitrary curvilinear 3D coordinates as a flat problem through the application of conformal transformations. This procedure requires finding a special coordinate system, known as \textit{isothermal coordinates} \cite{che55pam,lee18riem}, which is orthogonal and whose coordinate vectors have equal but non-unit lengths.

The membrane's complex geometry is described within the framework of differential geometry. The advantage of differential geometry is that it provides a well-established framework that enables a rigorous and compact description of complex geometries. The significance of differential geometry had been proven in fundamental theories such as General Relativity and String theory \cite{hob06cup}.

\subsection{Membrane parametrization}

The object of consideration is a curvilinear membrane of thickness $L_\text{d}$ whose modeling in 3D coordinates needs to be reduced to the isothermal coordinates in a plain domain. To this end, we begin with the Cartesian coordinates in 3D and consequently transit to a 2D description through a series of transformations and reductions. 

We begin with the notation and conventions and refer to Appendix \ref{sec:diff} for details on differential geometry. Namely, we use the indices $i, j, k = 1, 2, 3$ for the Cartesian coordinates $x^i = (x, y, z)$.
The metric tensor describing the entire 3D space is represented by the Kronecker delta
\begin{equation}\label{eq:flat_metric}
    dl^2_\text{3D} = \delta_{ij} dx^i dx^j = dx^2 + dy^2 + dz^2.
\end{equation}
The 3D vectors in the Cartesian coordinates are denoted by bold symbols, e.g., $r^i = \mathbf{r}$. Next, the curvilinear coordinates $\xi^a = (\xi, \eta, w)$ that suit the membrane description are indexed with $a, b, c = 1, 2, 3$ and the corresponding metric tensor is $g_{ab}$. When the transverse coordinate $w$ is omitted and one deals with the 2D curvilinear coordinate system $\xi^\mu = (\xi, \eta)$, we use the indices $\mu,\nu = 1, 2$, the metric tensor $\gamma_{\mu\nu}$, and other Greek symbols. In addition, if 2D coordinates are isothermal, the notation $u^\mu = (u, v)$ is used. Also, we use $\nabla_a$ for 3D covariant derivatives and $\mathcal{D}_\mu$ for 2D covariant derivatives. 

Next, we consider each coordinate system and the transformations between them in detail. The relations of the mappings between the different coordinate systems are illustrated in Fig.\,\ref{fig:1}.
Specifically, we consider a surface $\mathcal{S}$ formed by the points of the membrane in the middle of its thickness and given as
\begin{equation} \label{eq:surface}
    x^i = S^i(\xi^\mu),\qquad
\end{equation}
The unit normal vector $n^i(\xi^\mu)$ to the surface $\mathcal{S}$ satisfies the normality condition $n_i \partial_\mu S^i = 0$ and the unity condition $n^i n_i = 1$. By substituting Eq. (\ref{eq:surface}) into Eq. (\ref{eq:flat_metric}) and comparing with $dl^2_\text{2D}= \gamma_{\mu\nu}d\xi^\mu d\xi^\nu$, we obtain the induced metric tensor of the surface $S$
\begin{equation}
    \gamma_{\mu\nu} = 
    \delta_{ij}(\partial_\mu S^i) (\partial_\nu S^j).
\end{equation}
Also, we define the extrinsic curvature tensor $\chi_{\mu\nu}$
\begin{equation}
    \chi_{\mu\nu} =
    - n_{i} \partial_\nu\partial_\mu S^i.
\end{equation}
The trace and the determinant of the matrix ${K^\mu}_\nu = \gamma^{\mu\lambda}\chi_{\lambda\nu}$ represent the twice mean curvature $2M={K^\mu}_\mu=k_++k_-$ and the Gaussian curvature $K=\text{det}\,{K^\mu}_\nu=k_+k_-$, respectively. Here, $k_\pm$ are the principal curvatures calculated as the eigenvalues of ${K^\mu}_\lambda$.

\begin{figure}[t!]
    \centering    \includegraphics[width=0.8\linewidth]{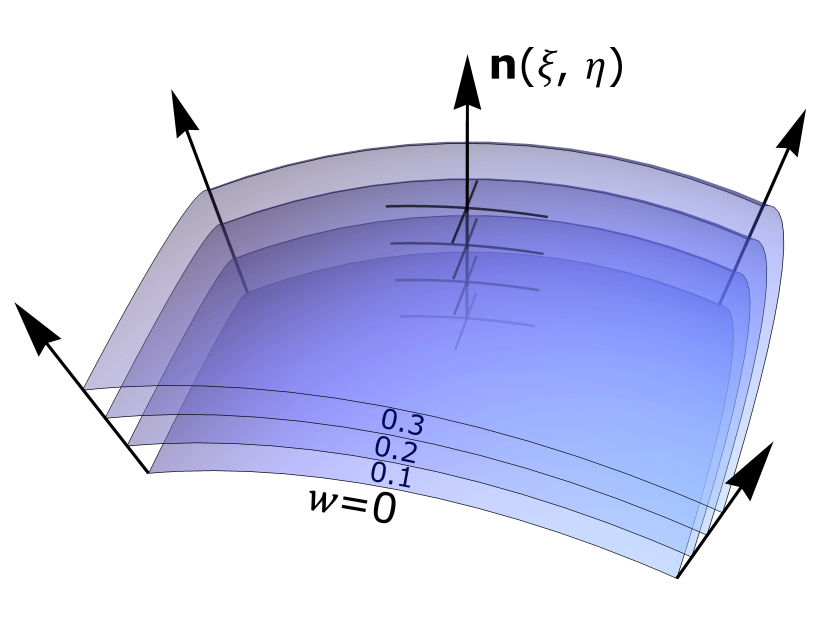}
    \caption{Surface evolution along the normal vector.}
    \label{fig:2}
\end{figure}

A 3D membrane can be described by adding a coordinate along the normal direction,
\begin{equation}\label{eq:transformation}
    x^{i}(\xi, \eta, w) =  S^i(\xi, \eta) + n^i(\xi, \eta) w,
\end{equation}
where $w\in[-L_\text{d}/2,L_\text{d}/2]$.
Accordingly, the new set of coordinates in the 3D space is $\xi^a = (\xi, \eta, w)$. Each value of $w$ generates its own surface $S_w$ which can be obtained by translating the points of the surface $S$ along the vector $n^i$, see Fig.\,\ref{fig:2}. 

Using the identity $(\partial_\mu n^i)(\partial_\nu n_i) = \chi_{\mu\alpha} \gamma^{\alpha\beta} \chi_{\beta\nu}$ and the relation $\gamma^{\alpha\beta} = \left(\gamma_{\alpha\beta}\right)^{-1}$, the substitution of Eq.\,\eqref{eq:transformation} into Eq.\,\eqref{eq:flat_metric} yields the metric tensor $g_{ab}$ for the 3D space in the coordinates $\xi^a$,
\begin{align}\label{eq:3d_metric}
    dl^2
    &= g_{ab} d\xi^a d\xi^b
    = {G_{\mu\nu}} d\xi^\mu d\xi^\nu
    + dw^2,
    \\\nonumber
    G_{\mu\nu} &\equiv
        \gamma_{\mu\nu}
        + 2 \chi_{\mu\nu} w
        + \chi_{\mu\alpha} \gamma^{\alpha\beta} \chi_{\beta\nu} w^2
    \\\nonumber
    &=(\delta_\mu^\alpha + w {K^\alpha}_\mu) \gamma_{\alpha\beta} (\delta_\nu^\beta + w {K^\beta}_\nu).
\end{align}
For Eq.\,\eqref{eq:3d_metric}, the metric tensor was split into two parts
\begin{equation}\label{eq:g_block}
    g_{ab} = \begin{pmatrix}
        G_{\mu\nu} & 0\\
        0 & 1
    \end{pmatrix},
    \qquad
    g^{ab} = \begin{pmatrix}
        G^{\mu\nu} & 0\\
        0 & 1
    \end{pmatrix},
\end{equation}
where $G^{\mu\nu}$ is defined as inverse to the matrix $G_{\mu\nu}$. The matrix $G_{\mu\nu}$ at some constant $w$ represents an induced metric of the surface defined by Eq. (\ref{eq:transformation}) at the corresponding value of $w$. Thus, $G_{\mu\nu}=\gamma_{\mu\nu}$ at $w=0$, as follows from Eq. (\ref{eq:3d_metric}). For compactness of notation, we will write for the determinants 
\begin{equation}
    g = \text{det}\,g_{ab},\qquad
    G = \text{det}\,G_{\mu\nu},\qquad
    \gamma = \text{det}\,\gamma_{\mu\nu},
\end{equation}
where, according to Eq. (\ref{eq:g_block}), the identity $g = G$ holds. Using Eq. (\ref{eq:3d_metric}), the explicit expression for $G$ reads
\begin{align}\label{eq:g_det}
    G &= \text{det}(G_{\nu\lambda}) = \gamma  
    \left(\text{det}(\delta_\mu^\alpha + w {K^\alpha}_\mu)\right)^2
    \\\nonumber &
    =\gamma (1+wk_+)^2(1+wk_-)^2
    = \gamma (1 + 2wM + w^2 K)^2.
\end{align}

\subsection{Conformally flat metric} 

According to the uniformization theorem \cite{riemann1851,poincare1908,koebe1907}, a generalization of the Riemann mapping theorem, any simply-connected 2D surface is conformally equivalent to one of the three Riemann surfaces: the open unit disk, the complex plane or the Riemann sphere. This means that any simply-connected surface can be parametrized by isothermal coordinates.

If some coordinate transformation from the initial 2D coordinates to the isothermal coordinates ${\xi^\mu = T^\mu(u^\nu)}$ is known, then the metric tensor is transformed as follows
\begin{align}\label{eE2sigma}
    \gamma_{\mu\nu} d\xi^\mu d\xi^\nu & =
    \gamma_{\mu\nu} (\partial_\alpha T^\mu)(\partial_\beta T^\nu) du^\alpha du^\beta
    \\\nonumber&
    =
    e^{2\sigma}\delta_{\mu\nu} du^\mu du^\nu.
\end{align}
In Eq.\,\eqref{eE2sigma}, $e^{2\sigma}$ is a conformal factor depending on the coordinates and the second line of Eq.\,\eqref{eE2sigma} follows from the condition that the coordinates $u^\mu$ are isothermal. The isothermal coordinates are named for their use in heat-equation problems as these coordinates have the physical meaning of constant-temperature lines in thermodynamics.

Details on how these transformations can be obtained numerically will be provided later. At this stage, it is important to note that the metric tensor corresponding to the isothermal coordinates has a conformally flat form 
\begin{equation}\label{eq:to_iso}
\gamma_{\mu\nu} \to \gamma'_{\mu\nu} = e^{2\sigma}\delta_{\mu\nu},
\end{equation}
which differs from a flat metric by the factor $e^{2\sigma}$.

\subsection{Operators} 

The next task is to express the 3D differential operators through the 2D differential operators using the formalism of differential geometry in the limit of a vanishing thickness. In the formalism of differential geometry, the divergence (div), the Laplace-Beltrami operator ($\Delta$), and the gauge-covariant Laplace-Beltrami operator ($\Delta_g$) read
\begin{subequations}
\label{eq:3d_operators}
\begin{align}
    \text{div}\,\mathbf{j} 
    &= \nabla_a j^a = \frac{1}{\sqrt{g}}\partial_a \left(\sqrt{g}g^{ab} j_b\right)
    \\
    \Delta \psi 
    &= g^{ab} \nabla_a \nabla_b \psi = \frac{1}{\sqrt{g}} \partial_a\left(\sqrt{g} g^{ab}\partial_b \psi\right)
    \\
    \Delta_\text{g} \psi 
    &= g^{ab} (\nabla_a - i A_a)(\nabla_b - i A_b) \psi
     \\\nonumber &
     =
    \Delta \psi
    - i (\nabla_a A^a) \psi
    - 2 i A^a \partial_a \psi
    - A_a A^a \psi,
\end{align}
\end{subequations}
where $A^a = g^{ab}A_b$ is a vector potential that corresponds to the magnetic field $\mathbf{B} = \text{curl} \mathbf{A}$, and $\mathbf{j}$ is a vector field. Accordingly, the 2D operators are defined as
\begin{subequations}
\label{eq:2d_operators}
\begin{align}
    \text{div}^\text{2D}\,\mathbf{j} 
    &= \mathcal{D}_\mu j^\mu = \frac{1}{\sqrt{\gamma}}\partial_\mu \left(\sqrt{\gamma}\gamma^{\mu \nu} j_\nu\right)
    \\
    \Delta^\text{2D} \psi 
    &= \gamma^{\mu \nu} \mathcal{D}_\mu \mathcal{D}_\nu \psi = \frac{1}{\sqrt{\gamma}} \partial_\mu\left(\sqrt{\gamma} \gamma^{\mu \nu}\partial_\nu \psi\right)
    \\
    \Delta_\text{g}^\text{2D} \psi 
    &= \gamma^{\mu \nu} (\mathcal{D}_\mu - i A_\mu)(\mathcal{D}_\nu - i A_\nu) \psi
     \\\nonumber &
     =
    \Delta^\text{2D} \psi
    - i (\mathcal{D}_\mu A^\mu) \psi
    - 2 i A^\mu \partial_\mu \psi
    - A_\mu A^\mu \psi.
\end{align}
\end{subequations}
The 2D covariant derivatives $\mathcal{D}_\mu$ depend on the Christoffel symbols based on the induced metric tensor $\gamma_{\mu\nu}$ in the same manner as the Christoffel symbols in $\nabla_a$ are based on $g_{ab}$ (see Appendix \ref{sec:diff} and References \cite{hob06cup,lee18riem}).

In the isothermal coordinates [Eq.\,\eqref{eq:to_iso}], the 2D operators
\begin{subequations}
\label{eq:2d_operators_iso}
\begin{align}
    \text{div}^\text{2D}\,\mathbf{j}
    &=
    \mathrm{e}^{-2\sigma} \delta^{\mu\nu} \partial_\mu j_\nu,
    \\
    \Delta^\text{2D} \psi 
    &=
     \mathrm{e}^{-2\sigma} \delta^{\mu\nu} \partial_\mu \partial_\nu \psi,
     \\\label{eq:glb_iso}
    \Delta_\text{g}^\text{2D} \psi 
    &=
    \mathrm{e}^{-2\sigma} \delta^{\mu\nu} \left(\partial_\mu - i A_\mu\right)
    \left(\partial_\nu - i A_\nu\right) \psi
\end{align}
\end{subequations}
look, up to the \textit{conformal factor} $e^{-2\sigma}$, like the respective operators in the flat 2D Cartesian coordinates. Note that the covectors $j_\mu$ and $A_\mu$ are used in Eq. (\ref{eq:2d_operators_iso}) instead of the vectors $j^\mu$ and $A^\mu$. These covectors are expressed in the basis of the isothermal coordinates rather than the Cartesian ones. Although there is no distinction between the vectors and covectors in the Cartesian coordinates, this equivalence does not apply to the isothermal coordinates. For instance, $\delta^{\mu\nu} \partial_\mu j_\nu$ cannot be written as $\partial_\mu j^\mu$, as an additional factor appears in $j^\mu = \mathrm{e}^{-2\sigma} j_\mu$ in accordance with the rule of raising indices.

\subsection{Dimensional reduction}

To express the 3D operators (\ref{eq:3d_operators}) through the 2D operators (\ref{eq:2d_operators}), we split the 3D operators into the transverse and surface terms
\begin{subequations}
\label{eq:operators_split}
\begin{align}
    \text{div}\,\mathbf{j}
    &=
    \frac{1}{\sqrt{G}}\partial_\mu \left(\sqrt{G} G^{\mu\nu} j_\nu\right)
    + \partial_w j^w
    + N j^w,
    \\
    \Delta \psi 
    &=
     \frac{1}{\sqrt{G}} \partial_\mu\left(\sqrt{G} G^{\mu\nu}\partial_\nu \psi\right)
     + \partial_w^2 \psi
     + N \partial_w \psi,
     \\
    \Delta_\text{g} \psi 
    &=
    \Delta \psi
    - i (\text{div}\,\mathbf{A}) \psi
    - 2 i A^\mu \partial_\mu \psi
    - A_\mu A^\mu \psi
    \\\nonumber&\;
    - 2 i A_w \partial_w \psi
    - A_w^2 \psi,
    \\ 
    N
    &=  \partial_w \left(\ln \sqrt{G}\right)
    = \frac{2M + 2w K}{1 + 2wM + w^2 K},
\end{align}
\end{subequations}
where the function $N$ associated with the surface curvature was introduced. Note that Eqs.\,(\ref{eq:operators_split}) are exact since neither limit nor approximation has been applied so far.

In general, there are two aspects of reducing the dimensionality of the system from 3D to 2D. The first one is to remove or ``integrate out'' the transverse degree of freedom of the system. The second aspect consists in neglecting the dependence of the geometry and external fields on the transverse coordinate $w$.

Regarding the integrating out the transverse degree of freedom, the major potential pitfalls appear in derivatives in the normal direction $\partial_w$
\cite{jen1971ann}. To illustrate this, consider an example of a quantum particle in a 1D potential well, where its wave function $\psi$ obeys the boundary conditions $\psi(w=\pm L_\text{d} / 2) = 0$. The groundstate wavefunction of such a particle is $\psi \sim \cos\left(\pi w / L_\text{d}\right)$. Thus, each derivative $\partial_w$ provides an additional factor $\pi/L_\text{d}$. Further, we will keep in mind that $\partial_w$ may have the order of $1/L_\text{d}$. Other derivatives $\partial_\mu$ are considered to not produce factors of the order $1/L_\text{d}$. Although it is a natural assumption, it should be carefully evaluated for each system. In general, the procedure of reducing the degrees of freedom depends on the equations of motion of the system, and it has already been analyzed for various systems \cite{da1981quantum,jen1971ann,streubel2016magnetism,gen22nat,tononi2023low,ferrari2008schrodinger,ortix2015quantum,belov2006operator,szameit2010geometric,entin2001spin,batz2008linear}. In a case study in what follows, we will show how to handle these terms properly, leaving them untouched at this step. 

As a part of the second aspect, the geometry has to be reduced by taking the limit of vanishing thickness of the membrane $L_\text{d}\to 0$, $w \to 0$. In this limit, the matrix $G_{\mu\nu}$ tends to the induced metric $\gamma_{\mu\nu}$, i.e.,
\begin{equation}
    G_{\mu\nu} \to \gamma_{\mu\nu},
    \qquad
    G^{\mu\nu} \to \gamma^{\mu\nu},
    \qquad
    \sqrt{G}\to\sqrt{\gamma}.
\end{equation}
Omitting linear and higher-order terms in $w$ in Eq. (\ref{eq:operators_split}), the operators acquire the form
\begin{subequations}
\label{eq:operators_reduced}
\begin{align} 
    \text{div}\,\mathbf{j}
    &\to
    \text{div}^\text{2D}\,\mathbf{j}
    + \partial_w j^w 
    + N j^w,
    \\
    \Delta \psi 
    &\to
     \Delta^\text{2D} \psi
     + \partial_w^2 \psi
     + N\partial_w \psi,
     \\
    \Delta_\text{g} \psi 
    &\to
    \Delta_\text{g}^\text{2D} \psi
    + (\partial_w - i A_w)^2 \psi
    + N(\partial_w - i A_w) \psi.
\end{align}
\end{subequations}

The reduction is justified when the thickness is small. The criterion for small thickness can be quantified as follows. The quantity $\gamma^{\mu\lambda}G_{\lambda\nu}$ tends to the identity matrix $\delta^\mu_\nu$ in the limit $w\to 0$. In light of Eq. (\ref{eq:3d_metric}), this implies that the elements of the matrix $L_\text{d} {K^\alpha}_\mu$ are much smaller than 1. By changing the basis of the matrix to its eigenvectors, the latter can be diagonalized with the elements $k_-$ and $k_+$. Thus, the criterion for small thickness reads $L_\text{d} k_\pm \ll 1$.

\subsection{Vector potential and gauge transformations} 

Even for a uniform magnetic field, the vector potential has a nontrivial form in the basis of the coordinates $(\xi, \eta, w)$ by virtue of the basis transformations (see Appendix \ref{sec:diff}) by Eq.\,(\ref{eq:general_basis_transformations})
\begin{equation}
    A_\xi = \frac{\partial x^i}{\partial\xi} A_i,\qquad
    A_\eta = \frac{\partial x^i}{\partial\eta} A_i,\qquad
    A_w = n^i A_i.
\end{equation}
Nevertheless, gauge degrees of freedom in the vector potential $A^\mu$ can help simplify the expressions for the operators. Thus, arbitrary gauge transformations read
\begin{equation}
    A_a \to A'_a = A_a - \partial_a \alpha,
\end{equation}
where $\alpha$ is some smooth differentiable function. Fortunately, normal components of the vector potential $A_w$ can be set to zero when the following gauge transformation is performed
\begin{align}
    &
    \alpha(\xi, \eta, w) = \int A_w(\xi, \eta, w) dw
    \approx\\\nonumber&\qquad\approx
    \alpha'(\xi, \eta) + w A_w|_{w=0} + \frac{w^2}{2} \partial_w A_w|_{w=0} + \mathcal{O}(w^3),
\end{align}
where $\alpha'(\xi, \eta)$ is an arbitrary function that emerges as a constant of integration. It represents the remaining gauge degrees of freedom. Since we are free to fix the gauge, we will set the function $\alpha'$ equal to zero. Then, the tangential and normal components of the transformed vector potential read
\begin{align}
    &
    A_\mu \to A'_\mu = A_\mu - \partial_\mu \alpha,
    \\\nonumber &
    A_w \to A'_w = A_w - \partial_w \alpha = 0.
\end{align}
The component $A'_w$ is thus identically zero together with its derivatives with respect to the normal direction (both partial $\partial_w$ and covariant $\nabla_w$ of any order), so the normal component of the vector potential can be omitted in the gauge-covariant Laplace-Beltrami operator.

Furthermore, the tangential magnetic field is encoded in the component $\partial_w A'_\mu$. When the tangential magnetic field is considered negligible for dynamics of the system, the dependence of $A'_\mu$ on $w$ can be omitted for the limit $L_\text{d}\to 0$. Since the gauge transformation function $\alpha$ tends to 0, the transformed tangential components $A'_\mu|_{w=0}$ coincide with the untransformed components $A_\mu|_{w=0}$. However, if the tangential magnetic field is not sufficiently small, the 2D description of the problem becomes problematic, and 3D grids should be used in the simulations.

\subsection{Finding isothermal coordinates}\label{sIsotherm}

Finding isothermal coordinates is not always possible analytically for an arbitrary geometry. In what follows, we will use conformal mappings to find the isothermal coordinates. Generally, a conformal transformation is a transformation that preserves angles locally \cite{lee18riem}. Conformal transformations act on the metric tensor by multiplying it with a non-zero scalar function, $\gamma_{\mu\nu} \to \gamma'_{\mu\nu} = \mathrm{e}^{2\sigma}\gamma_{\mu\nu}$. Thus, the Cartesian coordinates on the flat domain mapped back onto the surface represent the isothermal coordinates.

\begin{figure}[t!]
    \centering    \includegraphics[width=0.75\linewidth]{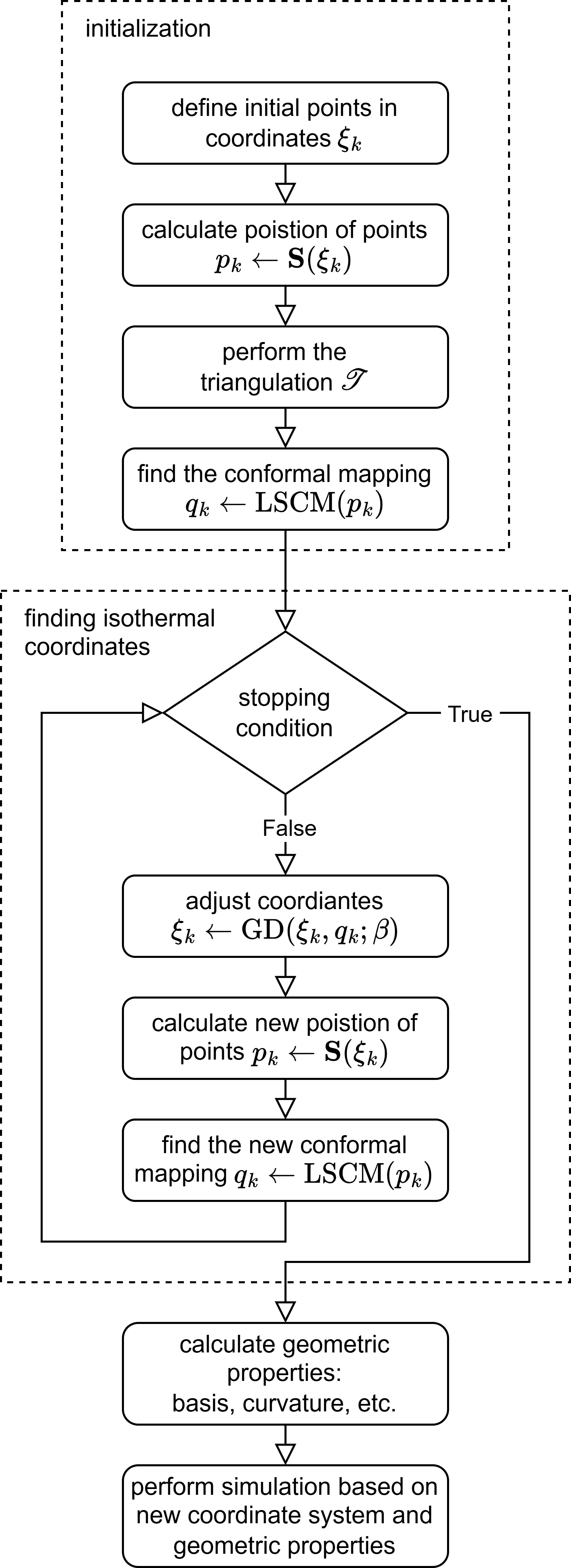}
    \caption{Block-diagram of the simulation algorithm with the preliminary finding of isothermal coordinates.}
    \label{fig:algorithm_diagram}
\end{figure}

Finding a conformal mapping of an arbitrary surface to a plain domain is a common problem in computer graphics, where it is solved by various algorithms\,\cite{meng2016tempo}. We choose a least-squares conformal mapping (LSCM) algorithm which was originally suggested \cite{levy2002least} as one of the possible options due to its simplicity, efficiency, and other technical advantages. In general, LSCM is a numerical algorithm that takes a triangulated surface with a finite number of points and computes the corresponding conformal mapping to a domain in a flat plane by optimizing a quadratic function \cite{levy2002least}. Our problem is the inverse one. Namely, a regular grid on a plane needs to be conformally mapped back to the surface. To this end, we modify the algorithm (see Fig.\,\ref{fig:algorithm_diagram}), applying it iteratively to find a uniform grid in the isothermal coordinates and adjusting the distribution of points on the surface using a gradient descent (GD).

In the block diagram in Fig.\,\ref{fig:algorithm_diagram}, $\xi_k$ are points in the $(\xi, \eta)$-coordinates, $p_k$ are points in the $(x, y, z)$-coordinates, $q_k$ are points in the $(u, v)$-coordinates, and $\textbf{S}$ is a mapping from the surface parametrization $(\xi,\eta)$ to the 3D-parametrization $(x, y, z)$. GD is a step of point adjustment with gradient descent. The parameter $\beta$ controls the speed of the optimization, which can vary for each iteration. 

The proposed algorithm for obtaining the isothermal coordinates is available as a code at \href{https://github.com/BogushPhysics/IsoCoord}{\textbf{IsoCoord}} \cite{repo}. In what follows we describe all the steps of the algorithm in detail.

\begin{figure}[ht]
    \centering    \includegraphics[width=0.7\linewidth]{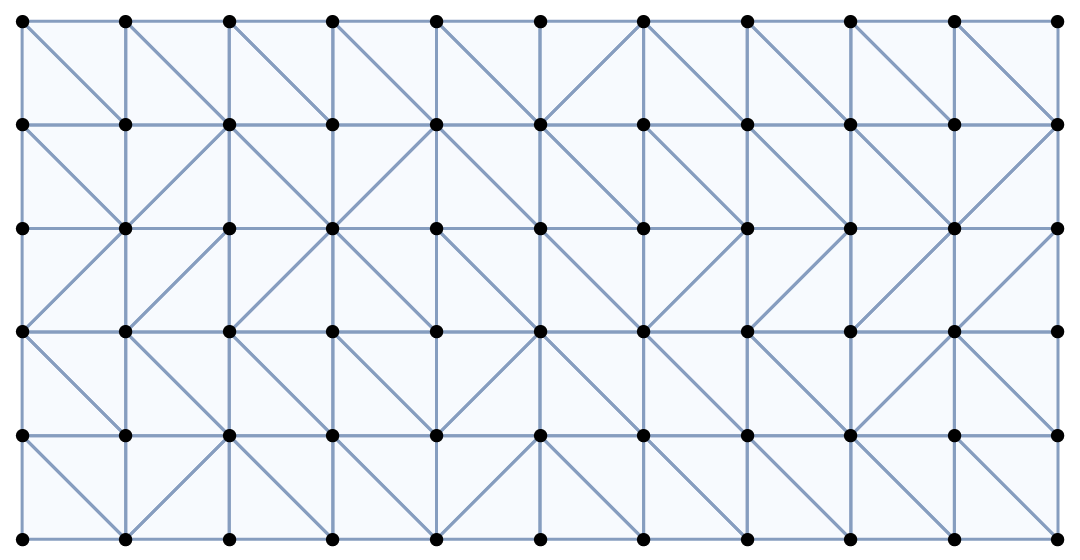}
    \caption{Triangulation of a regular grid.}
    \label{fig:4}
\end{figure}

\textbf{Point initialization.}
The points on the surface must be initialized such that they can be arranged into a rectangular array consistently. One way to initialize the points for a wide class of surfaces is to create an initially regular grid of points in the coordinates $(\xi, \eta)$. We assume that the triangulation of the points is already obtained by some known algorithm and the triangulation does not change during the iterations. For instance, the triangulation can be done using the Delaunay algorithm or by splitting each rectangle of the grid by a randomly chosen diagonal, see Fig.\,\ref{fig:4}.

\textbf{LSCM.} 
The LSCM is used to compute the conformal mapping $\mathcal{X}: (u, v) \mapsto (x, y, z)$ from the surface to a flat domain, i.e., from the isothermal coordinates $(u, v)$ to the points of the surface in a 3D Euclidean flat space $(x, y, z)\in S$. The main steps of the LSCM are outlined in Appendix \ref{sec:lscm}. Finding the conformal mapping $\mathcal{X}$ consists in the minimization of the following quadratic form
\begin{equation}\label{eq:quadratic}
    C(\mathcal{T}) = \mathbf{u}^\dagger \mathcal{M}^\dagger \mathcal{M} \mathbf{u},
\end{equation}
where $\mathbf{u}$ is a vector with complex values $\mathcal{U} = u + iv$ for all vertices in the triangulation $\mathcal{T}$, and $\mathcal{C}=\mathcal{M}^\dagger \mathcal{M}$ is a hermitian matrix corresponding to the quadratic form. As shown in Ref.\,\cite{levy2002least}, there are two degrees of freedom leaving the quadratic form (\ref{eq:quadratic}) invariant: a shift $\mathcal{U}_{j_i} \to \mathcal{U}_{j_i} + a$ and a rotation $\mathcal{U}_{j_i} \to b\, \mathcal{U}_{j_i}$, where $a$, $b$ are some complex constants. As a result, the matrix $\mathcal{C}$ is not full-rank. One way to remove these symmetry-related degrees of freedom was suggested in Ref. \cite{levy2002least} by fixing the values $\mathcal{U}$ at two points and splitting the system (\ref{eq:quadratic}) into homogeneous and inhomogeneous parts. In our approach, we similarly fix the points but formulate them as linear constraints. After that, the problem can be solved as a quadratic optimization with linear constraints, for example, by using numerical libraries such as \textbf{cvxopt} \cite{cvxopt}. The library reformulates the constrained quadratic optimization problem as an equivalent system of linear equations of the Karush-Kuhn-Tucker form
\begin{align}\label{eq:equivalent_linear}
    &
    \begin{pmatrix}
        \mathcal{C}_\mathbb{R} & \mathcal{A}^T \\
        \mathcal{A} & 0
    \end{pmatrix}
    \begin{pmatrix}
        \mathbf{u}_\mathbb{R} \\
        \lambda
    \end{pmatrix}
    =
    \begin{pmatrix}
        0 \\
        \mathbf{b}
    \end{pmatrix},
    \\\nonumber&
    \mathcal{C}_\mathbb{R} =
    \begin{pmatrix}
        \text{Re }\mathcal{C} & -\text{Im }\mathcal{C} \\
        \text{Im }\mathcal{C} & \text{Re }\mathcal{C}
    \end{pmatrix},\qquad
    \mathbf{u}_\mathbb{R} =
    \begin{pmatrix}
        \text{Re }\mathbf{u}\\
        \text{Im }\mathbf{u}
    \end{pmatrix}
\end{align}
where $\mathcal{C}_\mathbb{R}$ and $\mathbf{u}_\mathbb{R}$ are real-equivalent to $\mathcal{C}$ and $\mathbf{u}$, respectively, the linear constrain is encoded in $\mathcal{A} \mathbf{u}_\mathbb{R} = \mathbf{b}$, and $\lambda$ is a vector of auxiliary Lagrange multipliers. Then, the system of equations (\ref{eq:equivalent_linear}) is solved using the Cholesky decomposition. This approach preserves the performance of the algorithm and keeps the procedure of fixing points quite simple. Moreover, this allows for setting up more general constraints on the points, e.g., aligning two corner points at one horizontal line $v_i - v_j = 0$. To preserve the boundary of the membrane, we impose much stricter conditions on all boundary points, aligning them in a rectangle. However, in the case of stricter conditions, the system of linear equations does not admit an exact solution, and exact solvers used in cvxopt cannot be applied. Instead, we use the \textbf{scipy} library for sparse matrices to find an approximate solution by iterative methods such as the Quasi-Minimal Residual method \cite{scipy}. Finding the isothermal coordinates can be accelerated by applying the algorithm to the finer grids, interpolating the solution from the previous one.

\textbf{Point adjustment.}
The quadratic form given by Eq.\,\eqref{eq:quadratic} is quadratic in terms of the $(u,v)$-coordinates, but it is generally nonlinear in terms of the $(x,y,z)$-coordinates since the nonlinear surface equation constrains them. Thus, finding the gradient of the quadratic form with respect to the position of the points on the surface in terms of the $(x,y,z)$-coordinates can be complicated. Instead, we adjust the position of each point separately based on the local Jacobian matrix $\partial \xi^\mu / \partial u^\nu$. It corresponds to approximating the surface by tangent planes in the vicinity of each point.

In this step, we adjust the points of the grid on the surface such that they are aligned closer to a rectangular grid in the $(u,v)$-coordinates. The gradient descent step reads
\begin{subequations}\label{eq:gd}
\begin{equation}
    \eta^\text{(new)}_i = \eta_i + \beta \cdot \left(
          (u_i^* - u_i) \left.\frac{\partial \eta}{\partial u}\right|_i
        + (v_i^* - v_i) \left.\frac{\partial \eta}{\partial v}\right|_i
        \right),
\end{equation} 
\begin{equation}
    \xi^\text{(new)}_i = \xi_i + \beta \cdot \left(
          (u_i^* - u_i) \left.\frac{\partial \xi}{\partial u}\right|_i
        + (v_i^* - v_i) \left.\frac{\partial \xi}{\partial v}\right|_i
        \right).
\end{equation}
\end{subequations}
Here, the coordinates $u_i^*$ and $v_i^*$ are the target values of the $i$-th point in the rectangular grid. The derivatives $d\xi/du$, $d\xi/dv$, $d\eta/du$, $d\eta/dv$ are found numerically based on neighboring points in the grid. The constant $\beta$ controls the speed of the optimization. Too large values of $\beta$ may lead to divergence from the minimum while too small values of $\beta$ can result in a slow convergence.

\textbf{Stopping condition.}
The stopping condition can be different depending on the purpose. We use the maximal deviation from the target positions among the points $q_i$. If all points are not farther from the target positions than $\epsilon$, the searching for the isothermal coordinates stops. Other options include the use of mean values instead of the maximal ones or angle deviations from $90^\circ$ instead of position deviations.

\textbf{Geometric properties.}
After the isothermal coordinates are found, the corresponding conformal mapping can be used to calculate the geometric properties of the surface. The basis vectors and the normal vector can be found as 
\begin{equation}
    \mathrm{\mathbf{e}}_{u} = \frac{\partial \mathbf{r}}{\partial u},\qquad
    \mathrm{\mathbf{e}}_{v} = \frac{\partial \mathbf{r}}{\partial v},\qquad
    \mathbf{n} = \mathrm{\mathbf{e}}_{u} \times \mathrm{\mathbf{e}}_{v},
\end{equation}
where $\mathbf{r} = (x, y, z)$. The conformal factor can be calculated as a mean value of the squared norms of the basis vectors
\begin{equation}
    e^{2\sigma} = \frac{\mathrm{\mathbf{e}}_u^2 + \mathrm{\mathbf{e}}_v^2}{2}.
\end{equation}
Note that, in the case of an exact solution for the conformal mapping, the length of both vectors must be equal $|\mathrm{\mathbf{e}}_u| = |\mathrm{\mathbf{e}}_v|$.
However, since they are obtained numerically for a discrete grid, the lengths of the basis vectors can differ by a small value. The half-difference of the lengths relative to the mean length, 
\begin{equation}\label{eq:eta_accuracy}
\eta = (|\mathrm{\mathbf{e}}_u| - |\mathrm{\mathbf{e}}_v|) / (|\mathrm{\mathbf{e}}_u| + |\mathrm{\mathbf{e}}_v|),
\end{equation}
reflects the accuracy of numerically obtained isothermal coordinates.

The extrinsic curvature tensor is found as 
\begin{align}
    \chi_{\mu\nu} = - \frac{\partial^2 \mathbf{r}}{\partial u^\mu \partial u^\nu} \cdot \mathbf{n}.
\end{align}
The principal curvatures $k_\pm$ are the eigenvalues of the tensor ${K^{\mu}}_{\nu}=\gamma^{\mu\lambda}\chi_{\lambda\nu}$
\begin{equation}
    k_\pm = \frac{\chi_{uu} + \chi_{vv} \pm \kappa}{2e^{2\sigma}},\qquad
    \kappa^2 = (\chi_{uu} - \chi_{vv})^2 + 4 \chi_{uv}^2,
\end{equation}
which can be used for the mean $M$ and Gaussian $K$ curvatures,
\begin{align}
    &
    M = \frac{k_+ + k_-}{2} = \frac{\chi_{uu} + \chi_{vv}}{2e^{2\sigma}},
    \\\nonumber &
    K = k_+ k_- = \frac{\chi_{uu}\chi_{vv} - \chi_{uv}^2}{e^{2\sigma}}.
\end{align}

All aforementioned quantities can be calculated using central finite differences (except boundary values, which should be calculated with forward/backward difference) and do not require the calculation of derivatives of the order higher than two, which is plausible for the accuracy of the calculations.

\subsection{Applicability}
The applicability of the suggested algorithm depends on two groups of factors. The first group is related to the issue whether the found isothermal coordinates are suitable for numerical simulations using the FDM. The second group is related to the accuracy of the geometry approximation, $G_{\mu\nu}\approx\gamma_{\mu\nu}$.

The square root of the conformal factor $e^\sigma$ describes the distance between neighboring points. A more uniform conformal factor results in a more uniform distribution of points on the surface. To enhance the calculation accuracy, one should choose the grid resolution sufficiently fine at all points of the surface, especially where the conformal factor is maximal. If the conformal factor has a huge difference between the minimal and maximal values $\max(e^\sigma)/\min(e^\sigma)\gg1$, then either the calculations are not accurate enough in the domain of large $e^{\sigma}$ or the calculations are performed inefficiently in the domain of small $e^{\sigma}$. Thus, the ratio $\max(e^\sigma)/\min(e^\sigma)$ serves as a measure of how good performance and good accuracy can be achieved simultaneously in simulations. Also, instead of $\min(e^\sigma)$, one can use other statistical values of $e^\sigma$, such as its mean or median value over all points. If a conformal factor with a large max/min-ratio is inevitable, then multi-grid algorithms can be applied.

Although the membrane thickness is considered infinitesimal, the numerical calculation aims at simulating physical properties of a structure of finite thickness. The thickness of such a structure must be much smaller than the inverse of the principal curvatures $k_\pm$ of the membrane at every point. Otherwise, the approximation of $G_{\mu\nu}$ with the induced metric $\gamma_{\mu\nu}$ is no longer valid. Additionally, the grid step size must be much smaller than the inverse of the principal curvature $1/|k_\pm|$. Otherwise, the characteristic size of the geometric features is of the order of the grid size and the algorithm will lack accuracy.

The main purpose of the introduced method is to simulate thin 3D membranes using a 2D grid. However, the algorithm can also be applied to problems with 3D grids. In this case, the dynamic variable is not split into the transverse and surface degrees of freedom. Considering systems of a finite thickness may be relevant in the case of nonlinear systems, where the transverse and surface degrees of freedom cannot be split. Also, it can be relevant when the tangential magnetic field is estimated to be sufficiently strong to induce non-trivial transverse dynamics. In this case, the dependence of the vector potential $A_\mu$ on $w$ may not be omitted.

Finally, the fabrication of nanostructures with 3D geometries is associated with displacement of atoms and formation of defects in the crystal lattice. This may cause many additional effects affecting the system dynamics, such as a piezoelectric potential induced by shear strains and a strain-induced shift of the conduction band  \cite{Fomin2014}. Such effects are related to the lattice rather than the shape of the 3D geometry. If one considers physical systems where lattice effects play a crucial or appreciable role, corresponding strain-induced terms should be included in equations. In the present approach we restrict ourselves to the effects of the 3D geometry  only.

\section{Case study}

In this Section, we demonstrate capabilities of the developed method for some standard problem for a charged quantum particles in static electric and magnetic fields. Accordingly, as an example of the application of the developed method to stationary problems, we calculate the eigenstates and the spectrum of a charged particle in static electric and magnetic fields described by the Schr\"{o}dinger equation.

Specifically, we apply the algorithm to a quantum charged particle in magnetic and electric fields in a C-shaped geometry resembling a breached crater. First, we define a surface of this geometry by an analytical expression with two parameters. As two limiting cases, this expression contains a flat membrane and a ring crater without a breach. We demonstrate the finding of the isothermal coordinates for a breached crater and discuss its geometric properties. Next, the eigenstates and eigenvalues of the Schr\"{o}dinger equation solved using a 2D grid will be discussed. Particularly, the impact of the breach on the energy spectrum as a function of the magnetic field will be demonstrated for four examples: a flat membrane, a crater with a wide breach, a crater with a narrow breach, and a ring crater. However, 2D grids cannot capture the effects of strong tangential magnetic fields. Therefore, we will demonstrate how this method can be adapted for 3D grids to account for such effects.

\subsection{Geometry: C-shaped breached crater}

\begin{figure}[t]
    \centering    \includegraphics[width=1\linewidth]{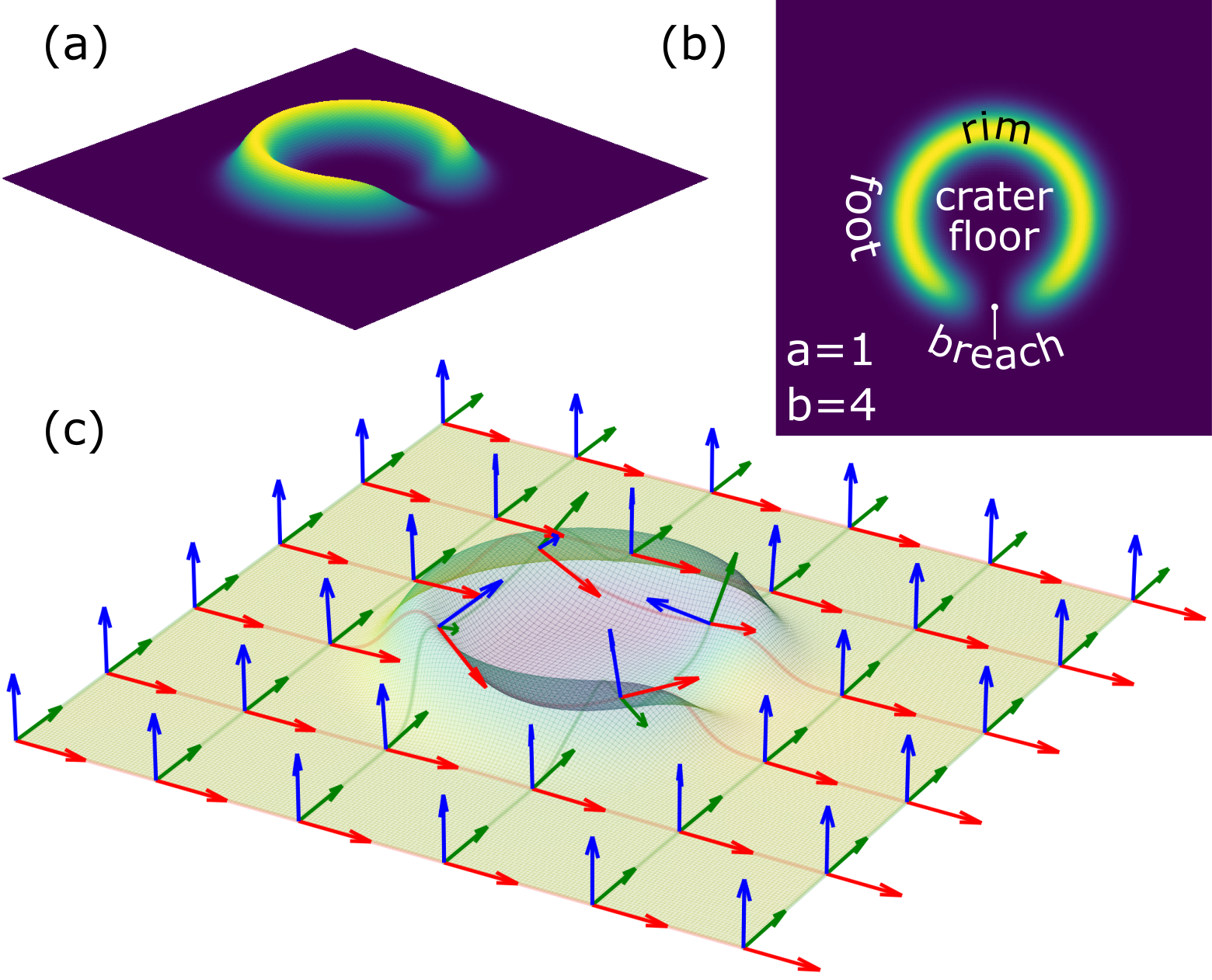}
    \caption{(a) Side and (b) top view of the C-shaped breached crater. (c) The local normalized basis and the normal vectors on the surface.}
    \label{fig:shape}
\end{figure}

We consider the surface of a C-shaped breached crater [see Fig.\,\ref{fig:shape}(a,b)] that is described by the following expression
\begin{equation}\label{eq:surface_example}
    z = e^{-4(r - 2)^2} (1 - 
    a e^{-b \theta^2}),
\end{equation}
where $r = \sqrt{x^2 + y^2}$, $\theta = \arctan(y, x)$ are the polar coordinates with $\theta\in(-\pi,\pi]$ and the dimensionless parameters $a, b$ determine the shape of the breach. Parameters are set to $a=1$, $b=4$ unless otherwise specified.

For the description of the shape, we will use the geology terminology used for breached craters as shown in Figure \ref{fig:shape}(b). The crater is surrounded by a rim. The outer part of the rim is the foot. The inner part encaged by the rim is the crater floor. The broken part of the rim is called the breach. We generated a 201$\times$201 grid in the range from $-5$ to $5$ in both $x$ and $y$ directions. With fewer than 100 iterations, the algorithm has produced a numerically calculated conformal mapping, yielding the isothermal coordinates. The thus obtained coordinate map is a square $(u, v)\in [0, W_u]\times[0, W_v]$ with $W_u=W_v=1$. The numerically calculated normalized local basis is shown in Fig.\,\ref{fig:shape}(c).

\begin{figure}[t]
    \centering    \includegraphics[width=1\linewidth]{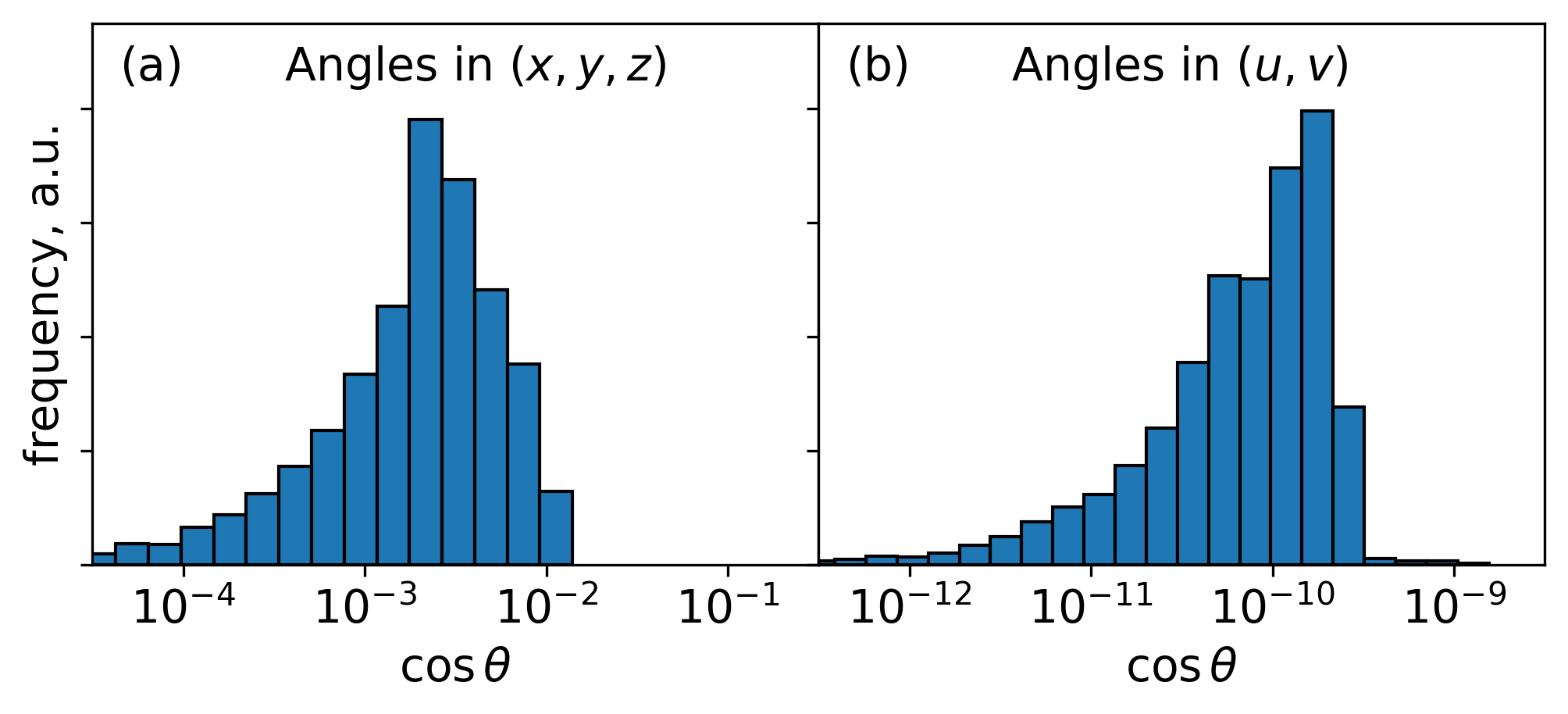}
    \caption{Distributions of the angle cosines between neighbor points in the grids of the 3D $(x,y,z)$-space (a) and the comformally mapped $(u, v)$-space (b).}
    \label{fig:angles}
\end{figure}
The angle distribution formed by the grid angles for the 3D and conformally mapped spaces is shown in Fig.\,\ref{fig:angles}. Ideally, all angles $\theta$ in $(u,v)$ should be $90^\circ$ yielding a cosine equal to zero. In fact, as shown in Fig. \ref{fig:angles}(b), the cosine of angles in the $(u, v)$-space differs from zero by a negligibly small value (from $10^{-9}$ to $10^{-14}$) due to the finite machine precision. The distribution of $\cos\theta$ in the $(x, y, z)$-space covers the range from $10^{-2}$ to $10^{-6}$ with a median around $10^{-3}$ in Fig.\,\ref{fig:angles}(a). The larger deviation from $90^\circ$ was anticipated as the grid points are not infinitesimally close to each other and the angle formed by three finitely separated points slightly differs from $90^\circ$. One can expect a smaller deviation from $90^\circ$ for finer grids. Also, we applied constraints on all boundary points resulting in an inexact solution of linear equations, slightly deforming the geometry. Instead, we could apply constraints only on four corner points, resulting in an exact solution with slightly deformed boundaries. The accuracy of the found isothermal coordinates can also be estimated by $\eta$ from Eq. (\ref{eq:eta_accuracy}), which does not exceed 1.5\%.

\begin{figure}[t]
    \centering    \includegraphics[width=1\linewidth]{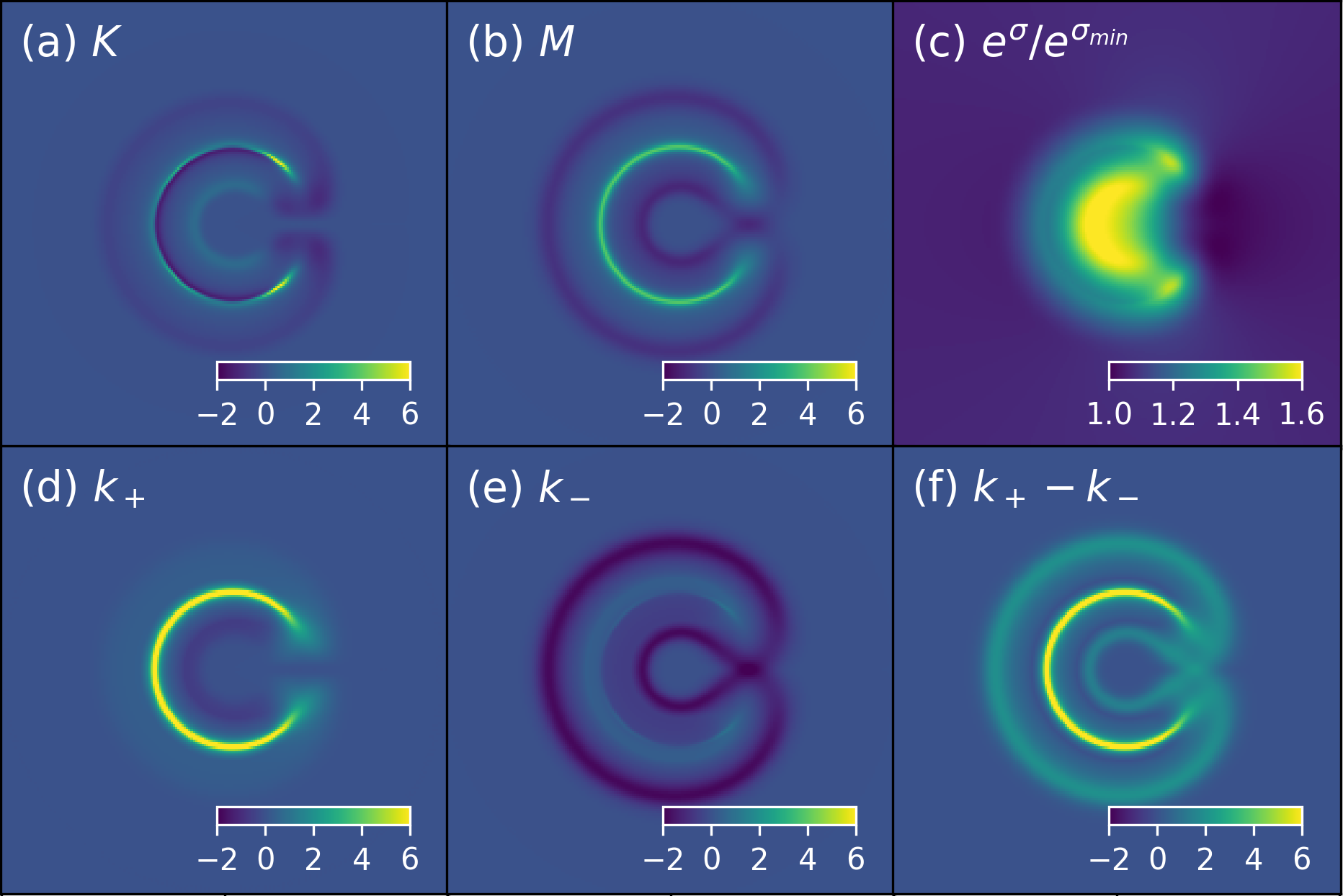}
    \caption{Distribution of the Gaussian curvature $K$ (a), mean curvature $M$ (b), square root of the conformal factor normalized to its minimal value $e^\sigma/e^{\sigma_\text{min}}$ (c), largest principal curvature $k_+$ (d),  smallest principal curvature $k_-$ (e), and the difference of the principal curvatures $k_+ - k_-$ (f). All quantities have been calculated for the C-shaped breached crater based on the numerically found conformal mapping.}
    \label{fig:curvatures}
\end{figure}

The geometric properties (square root of the normalized conformal factor and the curvatures) are shown in Fig. \ref{fig:curvatures}. The maximal value of the conformal factor is found for the floor (Fig. \ref{fig:curvatures}c), with $e^{\sigma_\text{max}}/e^{\sigma_\text{min}}\approx 1.7$. This means that the ratio of the smallest and the largest physical linear sizes of the grid step in the considered geometry of the breached crater is $\sim1.7$. The difference between the two principal curvatures, which will play an important role in the reduction of Schr\"{o}dinger equation, is practically zero almost everywhere except at two locations: At the top of the crater rim and at its foot. While at the foot, its value is $\sim$2, at the top it reaches $7.5$, that is, nearly four times larger. A potential well at the top of the rim has a purely geometrical nature.

\subsection{Schr\"{o}dinger equation in a 2D grid}

The stationary Schr\"{o}dinger equation of a charged particle in static electric and magnetic fields in a thin membrane reads
\begin{equation}
    -\Delta_g \psi + V_\text{n}(w)\psi + V_\text{t}(\xi, \eta)\psi = E \psi,
\end{equation}
where $V_\text{n}$ and $V_\text{t}$ are the transverse and surface potentials respectively. Some unspecified coordinate-independent consistent boundary conditions are considered at $w=\pm L_\text{d}/2$, where $L_\text{d}$ is the membrane thickness. The reduction of the stationary Schr\"{o}dinger equation from 3D to 2D was done in Refs. \cite{jen1971ann,da1981quantum}. Here, we briefly recall the steps to integrate out the transverse degrees of freedom.

After the reduction of the geometry from 3D to 2D, according to Eq. (\ref{eq:operators_reduced}), and omitting the normal components of the vector potential $A_w$, the Schr\"{o}dinger equation reads
\begin{equation}
    - \Delta_\text{g}^\text{2D} \psi
    - \partial_w^2 \psi
    - N \partial_w \psi
    + V_\text{n}(w)\psi + V_\text{t}(\xi, \eta)\psi = E \psi.
\end{equation}

One way to remove the transverse degrees of freedom is to separate variables, as suggested in Refs.\,\cite{jen1971ann,da1981quantum}. The wave function can be redefined as $\psi = \varphi / \sqrt{f}$, where $f = \sqrt{G/\gamma}$ can be found using Eq. (\ref{eq:g_det}), yielding
\begin{align}
    f = 1 + 2wM + w^2 K.
\end{align}
Then, one can rewrite the transverse part of the 3D Laplacian
\begin{align}
    \partial_w^2 \psi
     + N \partial_w \psi
     =
     \frac{1}{\sqrt{f}}\left(
         \partial_w^2 \varphi - \tilde{V}_\gamma \varphi
     \right),
\end{align}
where $\tilde{V}_\gamma$ is the geometric potential
\begin{equation}
    \tilde{V}_\gamma = \frac{\partial_w^2 \sqrt{f}}{\sqrt{f}}.
\end{equation}
Note that, apart from $\partial_w^2$, we transformed the term $N\partial_w \psi$ into the term $\tilde{V}_\gamma \varphi / \sqrt{f}$ which no longer contains derivatives with respect to $w$, and allows us to take the limit of the geometric potential
\begin{equation}
    V_\gamma = \lim_{w \to 0} \tilde{V}_\gamma = K - M^2 = -\frac{1}{4}(k_+ - k_-)^2.
\end{equation}
The resulting Schr\"{o}dinger equation reads
\begin{equation} \label{eq:sch_2}
    - \Delta_\text{g}^\text{2D} \varphi
    + V_\gamma \varphi
    - \partial_w^2 \varphi
    + V_\text{n}\varphi + V_\text{t}\varphi = E \varphi,
\end{equation}
where only the first two terms are approximated, while the others are exact.
The form of Eq. (\ref{eq:sch_2}) admits  separation of variables
$\varphi(\xi, \eta, w) = \varphi_\text{t}(\xi,\eta)\varphi_\text{n}(w)$,
leading to the independent transverse and surface equations
\begin{subequations}
\begin{align}
    &
    - \partial_w^2 \varphi_\text{n} + V_\text{n} \varphi_\text{n}
    = \lambda_\text{n} \varphi_\text{n},
    \\&\label{eq:schrodinger_reduced}
    -\Delta_\text{g}^\text{2D} \varphi_\text{t}
    + (V_\gamma + V_\text{t}) \varphi_\text{t}
    = (E - \lambda_\text{n}) \varphi_\text{t},
\end{align}
\end{subequations}
where $\lambda_\text{n}$ is the energy associated with the normal degree of freedom, which can be reabsorbed in the energy ${\tilde{E} = E - \lambda_n}$. For a small thickness $L_\text{d}$, only the ground state with the eigenvalue $\lambda_\text{0}$ can be considered since the energy difference between the two modes is of the order of $1/L_\text{d}^2$.

Similarly, a time-dependent Schr\"{o}dinger equation can be considered. For convenience, the wave function can be additionally multiplied by the global phase $e^{-i\lambda_\text{n} t}$ to remove the term $\lambda_\text{n}$. 

Substituting the 2D gauge-invariant Laplace-Beltrami operator in the isothermal coordinates (\ref{eq:2d_operators_iso}), the dimensionless stationary 2D Schr\"{o}dinger equation for a charged particle reads
\begin{equation}
    E \varphi = 
    -e^{-2\sigma}\delta^{\mu\nu}\left(\partial_\mu - i A_\mu\right)\left(\partial_\nu - i A_\nu\right) \varphi
    + \phi \varphi
    + V_\gamma \varphi,
\end{equation}
where $E$ and $\psi$ are the particle energy and wavefunction, respectively, $A_\mu$ is the magnetic vector-potential tangent to the surface, and $\phi$ is the electric scalar potential on the surface. The dimensional quantities can be restored by
\begin{equation}
    \bar{x} = L x,\quad  
    \bar{A} = \frac{\hbar}{qL} A,\quad
    \bar{\phi} = \frac{\hbar^2}{2m L^2q} \phi,\quad
    \bar{E} = \frac{\hbar^2}{2m L^2} E,
\end{equation}
where $L$ is some characteristic length, $m$ and $q$ are the mass and charge of the particle, respectively, and the dimensional quantities are denoted with overbars. The operator $\delta^{\mu\nu} (\partial_\mu - i A_\mu) (\partial_\nu - i A_\nu)$ is the flat gauge-covariant Laplace operator, which we discretize using the FDM with link variables. Here, we consider the Dirichlet boundary condition $\left.\varphi\right|_\text{boundary} = 0$, while the electric scalar potential and the magnetic vector potential are defined in Cartesian coordinates as
\begin{equation}\label{eq:potentials}
    \phi = -\mathcal{E} z,\qquad
    \mathbf{A} = \mathcal{B} y \, \mathbf{e}_{x}.
\end{equation}

\begin{figure*}[t]
    \centering    \includegraphics[width=1\linewidth]{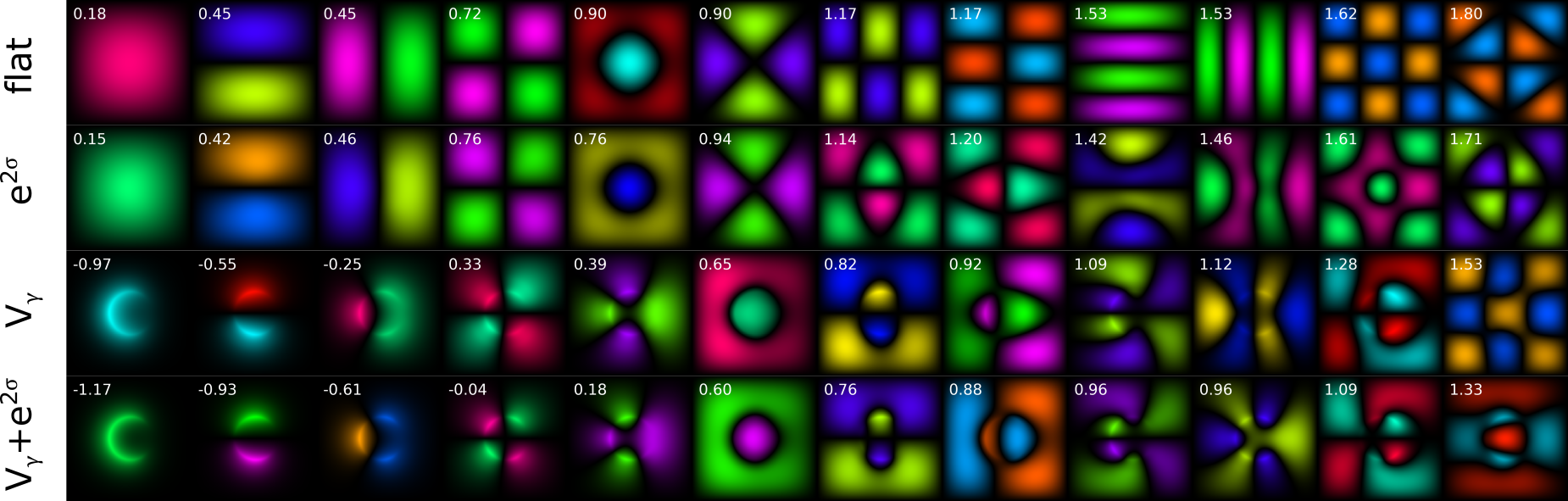}
    \caption{Eigenstates of the quantum particle without electric and magnetic fields modeled by four equations, from top to bottom: flat membrane, with conformal factor $e^{2\sigma}$, with geometric potential $V_\gamma$, with both conformal factor and geometric potential $V_\gamma+e^{2\sigma}$. In cases when the conformal factor function $e^{2\sigma}$ was not taken into account, it was replaced with its mean value $\left\langle e^{2\sigma}\right\rangle$. The energy of the eigenstates is shown in the left upper corner of each panel. The color intensity encodes the wave function amplitude $|\psi|$ and color encodes its phase.}
    \label{fig:eigenstates_comparison}
\end{figure*}

Figure \ref{fig:eigenstates_comparison} shows the wavefunctions and the energies of the eigenstates of the particle, thus allowing us to compare the effects of the curvature-induced term $V_\gamma$ and the factor $e^{2\sigma}$. Specifically, we consider the following four cases: the flat case (``flat''), only the conformal factor (``$e^{2\sigma}$''), only the geometric potential (``$V_\gamma$''), and both conformal factor and geometric potential (``$V_\gamma + e^{2\sigma}$''). In the cases without the conformal factor (``flat'' and ``$V_\gamma$''), we have replaced the conformal function $e^{2\sigma}$ with a constant equal to its average value $\left\langle e^{2\sigma}\right\rangle$ over all points. Then, the total area of the flat film and the curved membranes is the same, which allows for preserving the scale of the energy. In the flat case, some pairs of states are degenerate due to the symmetry of the square. The energy spectrum in the flat rectangular case is a textbook problem of quantum mechanics, with a known analytical solution
\begin{subequations}
\begin{equation}
    E^{(\text{flat})}_{n,m} = \pi^2 \left\langle e^{2\sigma}\right\rangle \left(
        \frac{n^2}{W_u^2} + \frac{m^2}{W_v^2}
    \right), 
\end{equation}
\begin{equation}
    \left|n,m\right\rangle = \frac{2}{\sqrt{W_u W_v}} \sin\left(\frac{\pi n u}{W_u}\right) \sin\left(\frac{\pi m v}{W_v}\right).
\end{equation}
\end{subequations}

The degenerate eigenstates are chosen such that they are similar to the eigenstates from the next row for $e^{2\sigma}$. The eigenstates of the Hamiltonian with included $e^{2\sigma}$ are very similar to those from the flat case. All of them resemble the distorted eigenstates of the flat case. In the case of the geometric potential ($V_\gamma$), the wave functions of the lower eigenstates are significantly changed, being localized near the minimum of $V_\gamma$. Some higher eigenstates from the row ``$V_\gamma$'' resemble those from the row ``$e^{2\sigma}$'', e.g. the eigenstates with the energies $0.65$ and $0.82$ in the ``$V_\gamma$'' row resemble the eigenstates with the energies $0.76$ and $1.14$ from the ``$e^{2\sigma}$'' row, respectively. In the case of all effects taken into account ($V_\gamma+e^{2\sigma}$), the eigenstates are similar to those that are calculated for the row ``$V_\gamma$'', however, the energy is appreciably shifted. All eigenstates in Fig.\,\ref{fig:eigenstates_comparison} are standing waves with zero probability current
\begin{equation}
    j_\mu = \Im\left(\psi^* (\partial_\mu - i A_\mu) \psi\right).
\end{equation}

\begin{figure*}[t]
    \centering    \includegraphics[width=1\linewidth]{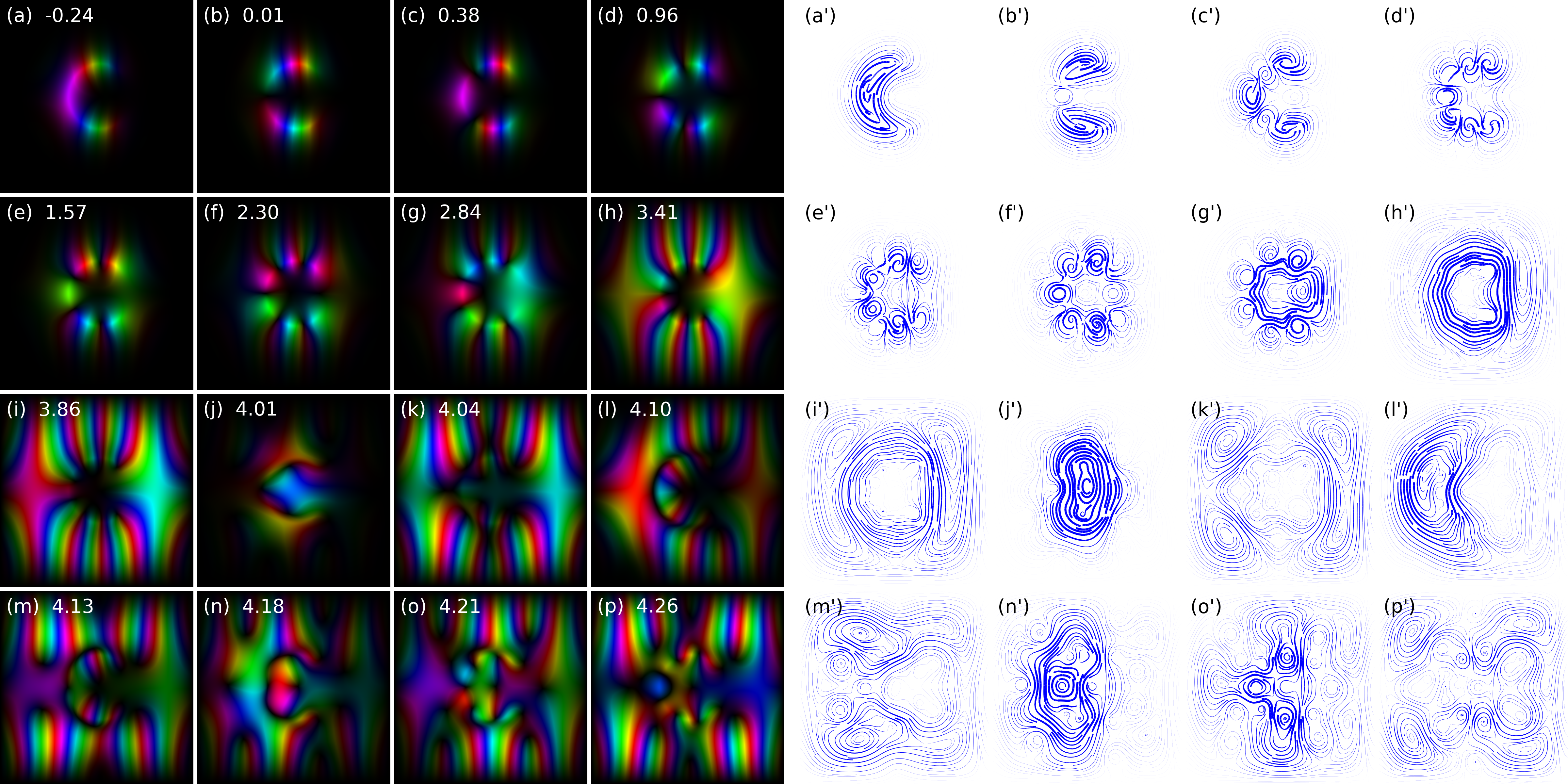}
    \caption{Wavefunctions (a-p) and probability currents (a'-p') for the first 16 eigenstates in a breached crater shape at $\mathcal{E} = 3$ and $\mathcal{B} = 1$.}
    \label{fig:eigenstates_currents}
\end{figure*}

The eigenstates calculated for homogeneous electric $\mathcal{E}=3$ and magnetic $\mathcal{B}=1$ fields directed along the $z$-axis are shown in Fig.\,\ref{fig:eigenstates_currents}(a-p). The nonzero normal magnetic field results in the stationary probability currents forming vortices shown in Fig.\,\ref{fig:eigenstates_currents}(a$'$-p$'$). The electric field can be used to push in or out the quantum particle from the top of the rim. In our case, we chose the electric field direction such that it additionally pushes the particle to the top of the rim. The lower modes (a-g) are localized on the top of the rim. The ground state (a) represents one contour of the current on the top of the rim and possesses a negative energy due to the effects of the negative geometric and electric potentials. Other lower states are similar to the ground state, but they have a number of nodes where the wave function is zero, resulting in $n$ contours of the current corresponding to the number of excitations. The state (j) is localized at the crater foot, while other states are not localized.

\begin{figure}[t]
    \centering    \includegraphics[width=1\linewidth]{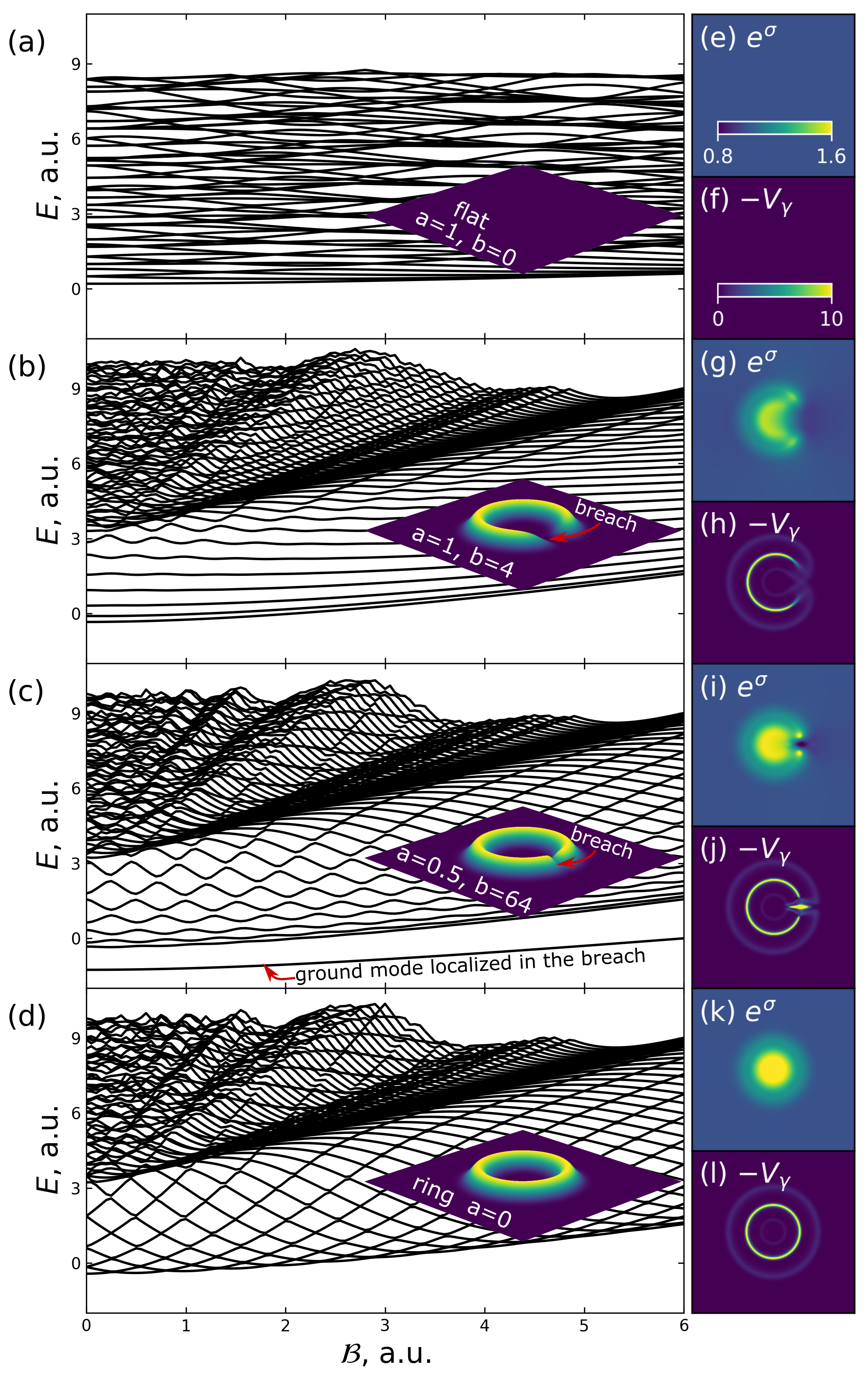}
    \caption{(a-d) Spectra of the energy of a quantum particle in an electric field with $\mathcal{E} = 3$ as a function of the magnetic field $\mathcal{B}$ for the first $60$ modes; (e,g,i,k) square root of the normalized conformal factor $e^\sigma$; (f,h,j,l) the geometric potential $V_\gamma$ (with inverse sign) for the following geometries: (a,e,f) flat square membrane, (b,g,h) breached crater with aforementioned parameters; (c,i,j) breached crater with a narrower breach; (d,k,l) ring-shaped crater. The parameters used for Eq. (\ref{eq:surface_example}) are indicated in the insets along with a visual representation of the geometries. The color scale in (e,f) applies to all contour plots.}
    \label{fig:spectrum}
\end{figure}

The evolution of the energy spectrum with increase of the magnetic field $\mathcal{B}$ is shown in Fig.\,\ref{fig:spectrum} for the following four geometries: a flat one, two C-shaped breached craters, and one crater with a ring rim. The spectrum for the flat rectangular membrane does not manifest a visible pattern in Fig.\,\ref{fig:spectrum}(a). The spectrum for the breached crater, Fig.\,\ref{fig:spectrum}(b), possesses lower modes with a large gap between the neighboring modes. These modes turn out to correspond to the eigenstates localized in the rim of the crater. Additionally, these modes exhibit a slightly oscillatory behavior as the magnetic field increases. For the breached crater with a smaller breach, Fig.\,\ref{fig:spectrum}(c), the spectrum reveals a similar pattern, but with a pronounced oscillatory behavior for the lower modes, leading to small gaps between two anticrossing modes. In addition, the energy of the ground state is appreciably lower than for the other modes. This ground mode corresponds to the state localized in the breach by virtue of a deep geometric potential well, Fig.\,\ref{fig:spectrum}(j). In the absence of a breach, the ring-shaped crater, Fig.\,\ref{fig:spectrum}(d), exhibits a similar spectrum with no detached mode localized in the crater breach. The similarity between the two latter spectra can be explained by the fact that the modes localized in the crater rim do not feel a small breach and behave like in a ring-shaped rim, while in Fig.\,\ref{fig:spectrum}(b) the breach is large and even does not create a potential well (compare Fig.\,\ref{fig:spectrum}(g) and Fig.\,\ref{fig:spectrum}(j)). The same effect was earlier proposed as an explanation for the Aharonov-Bohm effect in inhomogeneous Möbius rings \cite{Fomin2012} and for the oscillatory persistent currents in self-assembled simply-connected quantum craters \cite{Kleemans2007}.

\subsection{Ginzburg-Landau equation in a 3D grid}
\begin{figure}[t]
    \centering    \includegraphics[width=1\linewidth]{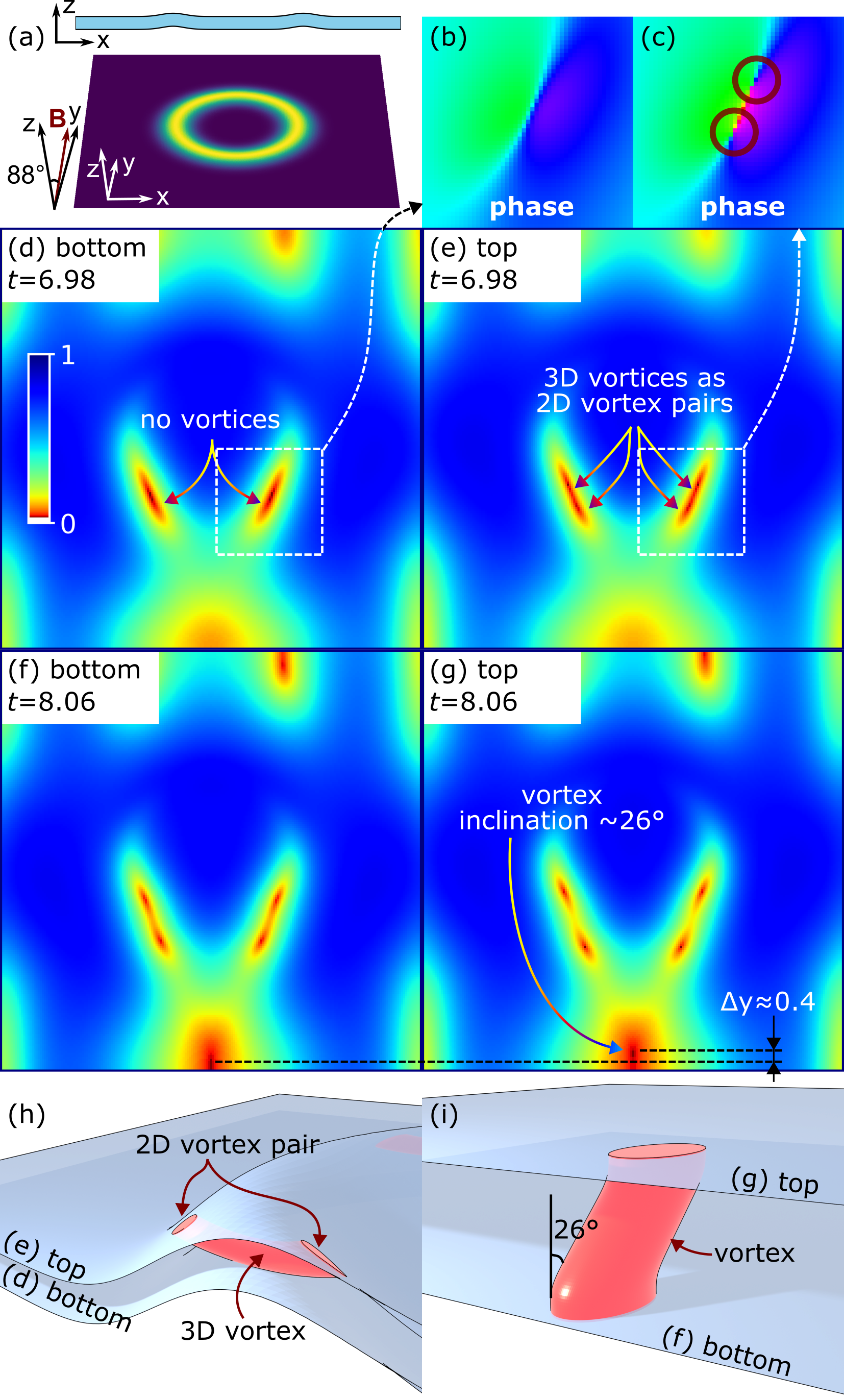}
    \caption{Snapshots of the magnitude of the superconducting order parameter $\left|\psi\right|$ obtained with a 3D grid for a ring-shaped crater in the magnetic field $\mathcal{B}=3$ inclined in the $(y,z)$-plane by $88^\circ$ with respect to the $z$-axis (d,e) at $t=6.98$ and (f,g) at $t=8.06$. The (d,f) bottom-most and (e,g) top-most surfaces are shown. The membrane geometry is presented in a cross-section at ${x=0}$ and a 3D schematic (a). The distributions of the order-parameter phase for the regions indicated in (d,e) are shown for the bottom-most (b) and top-most (c) surfaces. Schematic 3D representations of the vortex cores (h,i) correspond to the order parameter distributions in the rim for (d,e) and at the edge for (f,g), respectively.}
    \label{fig:vortices}
\end{figure}
For nonlinear equations, the surface and transverse dynamics usually cannot be split up exactly as it was for the Schr\"{o}dinger equation. Also, external fields can induce complicated transverse dynamics. In these cases, simulations with 2D grids can be less accurate than with 3D grids and can miss transverse effects. To demonstrate these effects, we consider a nonlinear time-dependent Ginzburg-Landau equation (TDGL). The TDGL equation describes the spatio-temporal evolution of the order parameter $\psi$ in a superconductor. The dimensionless TDGL equation in the presence of an external magnetic field reads
\begin{equation}
    \partial_t \psi = \Delta_\text{g} \psi + (1 - \left|\psi\right|^2)\psi.
\end{equation}
The boundary conditions imposed on the order parameter $\psi$ are $\bar{n}^a (\nabla_a - i A_a)\psi = 0$, where $\bar{n}^a$ is a normal vector to the boundaries of the membrane. The vector potential in Cartesian coordinates is ${\mathbf{A} = \mathcal{B} (y \cos\theta_B + z \sin \theta_B) \, \mathbf{e}_{x}}$.

For simplicity, the magnetic fields induced by the normal current and supercurrent are not considered. For example, this approximation is justified for Nb membranes with thicknesses $L_\text{d} \leq 50$ nm (see Ref.\,\cite{Bog24prb} for details on the applicability of a 2D TDGL approach). The dimensional units of length, time and magnetic field are $\xi$, $\xi^2/D$, and $H_\text{c2}=\Phi_0/2\pi\xi^2$, respectively, where $\xi$ is the coherence length, $D$ the diffusion coefficient, and $\Phi_0 \approx 2.07\times10^{-15}$ Wb the magnetic flux quantum. For instance, for moderately clean superconductor Nb films at $T/T_\mathrm{c} = 0.952$, these parameters are $60$\,nm, $2.8$\,ps, and $92$\,mT\,\cite{Dob12tsf} and the critical temperature $T_\text{c}\approx 9.2$\,K.

The TDGL admits the existence of topological defects, such as vortices. Each vortex has a core where the magnitude of the order parameter is small and turns to zero at the centerline of the core. A vortex is characterized by an integer number, vorticity, defined as a circulation of the phase gradient $\frac{1}{2\pi}\oint_\Gamma \partial_i \text{arg}\psi \,dl^i$ along some contour $\Gamma$ around the given vortex. It is worth mentioning that the contour orientation differs for the 3D and 2D representations. In the 3D representation, the vortex vorticity is defined with respect to the direction of the external magnetic field. In the 2D representation, the vorticity is defined with respect to the surface normal. Vortices with negative vorticity are called antivortices. In 3D nanoarchitectures, vortex core lines can be inclined or bent inside the walls \cite{smi20ltp,Mem24arx}, which also affects the vortex motion.

The geometry considered for the TDGL modeling is a ring-shaped crater described by the equation $z = 0.2\, e^{-(r - 4)^2}$ in the patch of the size $10\times 10$ and thickness 0.8 (i.e., ${w\in[-0.4,0.4]}$), as shown in Fig.\,\ref{fig:vortices}(a). The size corresponds to $600\,\text{nm}\times600\,\text{nm}\times48\,\text{nm}$ for Nb at $T/T_\text{c}=0.952$. The time step is set to 0.002 in dimensionless units, while the grid size is $201\times201\times6$ points. The magnetic field with strength $\mathcal{B}=3$ (which corresponds to 276 mT for Nb at $T/T_\text{c}=0.952$) is tilted at an angle $\theta_B = 88^\circ$ with respect to the axis $z$ in the $(y,z)$-plane. Then, the magnetic field is almost tangential at all points of the membrane, except the crater's rim. The field tilt breaks the local symmetry near the edge of the membrane. As a result, a vortex pierces the membrane at some angle with respect to the membrane normal. 

Snapshots of $|\psi|$ at two specific time instants are shown in Fig.\,\ref{fig:vortices}(d-g). In these snapshots, the first vortex nucleations are shown rather than an equilibrium state with vortices in a stable state. In Figs.\,\ref{fig:vortices}(d,e), the order parameter magnitude is shown for the bottom-most ($w=-0.4$) and the top-most ($w=0.4$) surfaces at the time instant $t=6.98$. The vortices nucleate not at the edge of the membrane, but at the rim of the crater. On the bottom-most surface of the membrane, there are no vortices at all, Fig.\,\ref{fig:vortices}(d), while there are two 2D vortex-antivortex pairs on the top-most surface, Fig.\,\ref{fig:vortices}(e). Distributions of the order-parameter phase are shown in Figs. \ref{fig:vortices}(b) and (c) to augment the interpretation. Each vortex-antivortex pair in the 2D representation as shown in Fig.\,\ref{fig:vortices}(h) corresponds to a single vortex in the 3D representation. Accordingly, a single vortex in the 3D representation corresponds to a vortex-antivortex pair in the 2D representation if the direction of the magnetic field flips with respect to the normal vector of the membrane. At a later time instant $t=8.06$, Figs.\,\ref{fig:vortices}(f,g), the vortices are nucleating at the edges. As shown by two dashed lines, the center of the vortex core on the top-most surface is not strictly above its center on the bottom-most surface. This results in an inclination of the vortex core with respect to the surface normal by $\sim 26^\circ$ as shown in Fig.\,\ref{fig:vortices}(i).

To summarize, the features of the vortex dynamics revealed in this simulation can be accurately captured using 3D grids only, while 2D grids implicitly assume that transverse dynamics do not significantly affect the overall behavior. The developed algorithm therefore offers the flexibility to select a 3D grid for modeling membranes in a magnetic field with a significant tangential component or a 2D grid for enhanced performance whenever dynamics of transverse degrees of freedom can be neglected. 

\section{Conclusion}

We have developed a method to perform simulations of physical systems in curved 2D surfaces of an infinitely small thickness within the finite difference paradigm. This has been achieved with a preceding application of a conformal transformation to the initial geometry of the surface, resulting in a parametrization of the surface in terms of the isothermal coordinates. The performed conformal mapping allows one to reformulate differential operators involved in simulations in terms of differential operators of a flat geometry at the cost of adding algebraic terms. The introduced procedure enlarges the palette of possible membrane geometries that can be used for simulations with finite difference methods. Though we have suggested an iterative LSCM algorithm for finding conformal mappings when obtaining isothermal coordinates, there are a plethora of methods for finding conformal mappings that can be used for the purpose of simulations of physical systems in 3D membranes within the suggested scheme.

In case when transverse degrees of freedom cannot be neglected due to nonlinearity or strong external fields, the proposed algorithm can be formulated for 3D grids with a few points along the normal direction. This is achieved by omitting the linear and higher-order terms in thickness in the geometry-related terms in the equations of motion for the physical system under study. This extension can be particularly valuable for such equations as the TDGL equation, which requires accounting for possible variation of the order parameter in the normal direction and the impact of a strong tangential magnetic field. 

The proposed algorithm translates the calculations of geometric properties of curved surfaces to the numerical framework. That is, there is no need to manually calculate and analyze terms specific to each new geometry, including custom contributions from the differential operators in curved systems. Instead, the algorithm incorporates all geometric information through the multiplicative conformal factor and the curvature terms. The differential operator transformations under conformal mappings are also well-studied in differential geometry. The adherence to the finite difference scheme paradigm should enable researchers to apply their existing developments to curved geometries with minimal code modifications and reduced risk of performance deterioration. All these features can simplify the simulation of various physical systems on thin curved membranes, establishing a solid application domain for the proposed approach.

The developed algorithm for obtaining the isothermal coordinates is made freely available at \href{https://github.com/BogushPhysics/IsoCoord}{\textbf{IsoCoord}} \cite{repo}. The proposed paradigm sets a stage for advancing the simulations of physical properties of 3D membranes across various fields of condensed matter physics, materials science, and electronics, ranging from superconductivity and fluxonics over magnetism and spintronics to optics and acoustics.

\acknowledgments
Work by I.B. was funded by the Deutsche Forschungsgemeinschaft (DFG, German Research Foundation) under Germany’s Excellence Strategy -- EXC-2123 QuantumFrontiers -- 390837967. The research is based upon work from COST Action CA21144 (SuperQuMap) supported by E-COST (European Cooperation in Science and Technology).

\appendix
\renewcommand{\theequation}{\thesection\arabic{equation}}
\section{Differential geometry}\label{sec:diff}

In a physical context, an introduction to differential geometry can be found in Ref.\,\cite{hob06cup}. Here, we just briefly outline the essentials of differential geometry relevant for Sec.\,\ref{sModel}.

Let a manifold $\mathcal{M}$ of dimension $D$ represent some space, such as a 3D space or a 2D surface. The manifold is parametrized by a set of coordinates $q^i=(q^1, q^2, \ldots)$, with the index $i$ running from 1 to $D$. At every point $p$ of the manifold $\mathcal{M}$, one can define vectors $v^i$ as elements of the tangent space $T_p\mathcal{M}$ and covectors $v_i$ as elements of the cotangent space $T^*_p\mathcal{M}$. An arbitrary tensor can have both vectorial and covectorial indices, $v^{\;\,j\;\;l}_{i\;\;k}$. Physically, the tangent and co-tangent spaces are dual to each other, like the space of velocities and the space of momenta in the Hamiltonian mechanics. 

The central object of differential geometry is the metric tensor $g_{ij}$ which defines an infinitesimal line element $dl$
\begin{equation}
    dl^2 = g_{ij} dx^i dx^j,
\end{equation}
where $dx^i$ is a differential of the coordinate. Here, we use the notation of Einstein summation, implying that there is a summation by repeating indices over all their values, $v_a v^a = v_1 v^1 + \ldots + v_D v^D$. The inverse metric tensor $g^{ij}$ is defined by the inverse matrix of the metric tensor, $g_{ij}g^{jk} = \delta_{i}^{k}$, with $\delta_{i}^{k}$ being the Kronecker delta. Vectors and covectors can be translated to each other by raising and lowering indices, $v_i = g_{ij} v^j$ and $v^i = g^{ij} v_j$. The squared length of a vector or covector is defined by the (inverse) metric tensor: $|v|^2 = g_{ij} v^i v^j = g^{ij} v_i v_j$. 

In addition to the metric tensor, there is a totally antisymmetric Levi-Civita tensor, which is used for the curl operator,
\begin{equation}
    \epsilon_{i_1 i_2 \ldots i_D} = \pm \sqrt{g},\qquad
    \epsilon^{i_1 i_2 \ldots i_D} = \pm 1 / \sqrt{g},
\end{equation}
where $g = \text{det}\,g_{ij}$. Thus, for $D=3$ one has $\epsilon_{123} = -\epsilon_{213} = \sqrt{g}$.

Differential geometry operates with covariant quantities only, and there are only three main ways to produce a covariant quantity: by tensorial multiplication ($v_{ij} = u_i w_j$), by contraction ($v_i = u_{ij} w^j$), or by acting with a covariant derivative ($v_{ij} = \nabla_i u_j$). The covariant derivative is defined as 
\begin{equation}
    \nabla_i = \partial_i + \begin{pmatrix}
        \text{algebraic}\\\text{terms}
    \end{pmatrix},
\end{equation}
where $\partial_i$ is a partial derivative with respect to the coordinate $q^i$ and the algebraic terms depend on the object the covariant derivative is acting on. If the object under differentiation is a tensorial quantity, the algebraic terms are expressed through the Christoffel symbol  $\Gamma^i_{jk} = \frac{1}{2} g^{il}\left(
        \partial_k g_{jl} + \partial_j g_{kl} - \partial_l g_{jk}
    \right)$,
\begin{subequations}
\begin{align}
    &\text{scalar:}
    &
    \nabla_i \psi &
    = \partial_i \psi,
    \\
    &\text{vector:}&
    \nabla_i v^j &= \partial_i v^j + \Gamma^j_{ik} v^k,
    \\
    &\text{covector:} &
    \nabla_i v_j &= \partial_i v_j - \Gamma^k_{ij} v_k,
    \\
    &\text{tensor:}&
    \nabla_i {v_{j}}^{k} &= \partial_i {v_{j}}^{k}
    - \Gamma^l_{ij} {v_{l}}^{k}
    + \Gamma^k_{il} {v_{j}}^{l},
\end{align}
\end{subequations}
The Christoffel symbols are defined in such a way that the metric tensor is covariantly constant, $\nabla_i g_{jk} = 0$, that justifies $g^{ab}\nabla_c v_a = \nabla_c (g^{ab} v_a) = \nabla_c v^b$.

Under a coordinate transformation from the coordinates $q^i$ to some coordinates $q'^a$, vectors and covectors are transformed according to the rules
\begin{align}\label{eRules}
    v^i \to v^{a} = {\mathrm{e}^a}_{i} v^i,\qquad
    v_i \to v_{a} = {\mathrm{e}^i}_{a} v_i.
\end{align}
In Eq.\,\eqref{eRules}, the transformation matrices
\begin{equation}\label{eq:general_basis_transformations}
    {\mathrm{e}^a}_{i} = \frac{\partial q'^{a}}{\partial q^i},
    \qquad
    {\mathrm{e}^i}_{a} = \frac{\partial q^i}{\partial q'^{a}},
\end{equation}
are inverse to each other, ${\mathrm{e}^a}_{i} {\mathrm{e}^i}_{b} = \delta^a_b$ and ${\mathrm{e}^i}_{a}{\mathrm{e}^a}_{j} = \delta^i_j$.

The above conventions are used throughout Sec.\,\ref{sModel}.

\section{Least square conformal mapping}
\label{sec:lscm}
For a conformal mapping $\mathcal{X}: (u, v) \mapsto (x, y, z)$ from the isothermal coordinates $(u, v)$ to the points of a surface in the 3D Euclidean flat space $(x, y, z)\in S$ the local basis is given by
\begin{equation}\label{eq:uv_basis}
    \mathrm{\mathbf{e}}_u = \frac{\partial\mathcal{X}}{\partial u},\qquad
    \mathrm{\mathbf{e}}_v = \frac{\partial\mathcal{X}}{\partial v},
\end{equation}
and the normal vector to the surface $\mathbf{n}$ is defined as
\begin{equation}\label{eq:uv_normal}
    \mathbf{n} = \frac{\mathrm{\mathbf{e}}_u \times \mathrm{\mathbf{e}}_v}{\left|\mathrm{\mathbf{e}}_u \times \mathrm{\mathbf{e}}_v\right|}.
\end{equation}
Since $\mathrm{\mathbf{e}}_u$ and $\mathrm{\mathbf{e}}_v$ are orthogonal and have the same norm, as required by the isothermal coordinates, the condition ${\mathbf{n} \times \mathrm{\mathbf{e}}_u = \mathrm{\mathbf{e}}_v}$ is satisfied \cite{levy2002least}.
After the application of a triangulation algorithm, the surface is represented by a set $\mathcal{T}$ of triangles consisting of triangles $\mathcal{T}_i$. Each triangle $\mathcal{T}_i$ is provided with a local orthonormal basis, with $(x_1, y_1), (x_2, y_2), (x_3, y_3)$ being the coordinates of its vertices in this basis. In particular, it is convenient to choose $x_1$, $x_2$, $y_1$ to be equal to 0. The local basis of all triangles must be oriented consistently to each other. 

We restrict the consideration of the mapping $\mathcal{X}$ to a triangle $\mathcal{T}_i$. Defining the inverse mapping $\mathcal{U}: (x, y) \mapsto (u, v)$, one obtains for $\mathcal{U} = u + i v$ the Cauchy-Riemann equation
\begin{equation}
    \frac{\partial\mathcal{U}}{\partial x} + i\frac{\partial\mathcal{U}}{\partial y} = 0,
\end{equation}
The criterion function to be optimized for the triangle $\mathcal{T}_i$ reads
\begin{equation}
    C(T_i) = \int_{\mathcal{T}_i} \left|\frac{\partial\mathcal{U}}{\partial x} + i\frac{\partial\mathcal{U}}{\partial y}\right|^2dA \approx 
    \left|\frac{\partial\mathcal{U}}{\partial x} + i\frac{\partial\mathcal{U}}{\partial y}\right|^2 A_{\mathcal{T}_i},
\end{equation}
where $A_{\mathcal{T}_i}$ is the area of the triangle $\mathcal{T}_i$. 

The full optimization procedure consists in the finding of the minimum of $C(T_i)$ for the entire triangulation
\begin{equation}
    C(\mathcal{T}) = \sum_{\mathcal{T}_i\in\mathcal{T}} C(\mathcal{T}_i).
\end{equation}
As shown in Ref.\,\cite{levy2002least}, up to higher-order corrections, the criterion function $C(\mathcal{T}_i)$ for a triangle can be presented as
\begin{equation}
    C(\mathcal{T}_j) = \frac{1}{A_{\mathcal{T}_j}}\left|
    (
        \mathcal{U}_{j_1},
        \mathcal{U}_{j_2},
        \mathcal{U}_{j_3}
    )
    \begin{pmatrix}
        \mathcal{X}_{j_3} - \mathcal{X}_{j_2}\\
        \mathcal{X}_{j_1} - \mathcal{X}_{j_3}\\
        \mathcal{X}_{j_2} - \mathcal{X}_{j_1}\\
    \end{pmatrix}
    \right|^2,
\end{equation} 
where $\mathcal{U}_{j_i}$ is the value of $\mathcal{U}$ at point $j_i$ of the triangle $T_i$. Similarly, $\mathcal{X}_{j_i}$ is the value of $x_{j_i} + i y_{j_i}$ at the point $j_i$ in the local basis of the triangle $j$. Taken together, this yields the quadratic form
\begin{equation}
    C(\mathcal{T}) = \mathbf{u}^\dagger \mathcal{M}^\dagger \mathcal{M} \mathbf{u},
\end{equation}
where $\mathbf{u}$ is a vector with complex values $\mathcal{U} = u + iv$ for all vertices in the triangulation $\mathcal{T}$ and $\mathcal{C}=\mathcal{M}^\dagger \mathcal{M}$ is a hermitian matrix corresponding to the quadratic form. Minimization of the quadratic form allows for finding the isothermal coordinates. This is achieved by applying the algorithm detailed in Sec.\,\ref{sIsotherm}.

\bibliography{main}

\end{document}